%% file: main.tex
\documentclass[12pt]{article}

\input{definitions.tex}
\usepackage{isotope}

\title{Snowmass2021 Cosmic Frontier White Paper: \\ Calibrations and backgrounds for dark matter direct detection
}

\date{\today}

\input{author_list.tex}

\begin{document}

\maketitle

\begin{abstract}
Future dark matter direct detection experiments will reach unprecedented levels of sensitivity.
Achieving this sensitivity will require more precise models of signal and background rates. 
Improving the precision of signal and background modeling goes hand-in-hand with novel calibration techniques that can probe rare processes and lower threshold detector response. 
The goal of this white paper is to outline community needs to meet the background and calibration requirements of next-generation dark matter direct detection experiments.
\end{abstract}

\tableofcontents

\pagebreak

\input{Sections/1-Introduction.tex}

\section{Calibration}
\label{sec:calibration}

Traditionally, detector response to scattering events has been divided into two types: nuclear recoil (NR) and electron recoil (ER). Detector calibration has typically followed the same division, with neutrons as a probe for NR calibration and with $\gamma$s or $\beta$s as a probe for ER calibration. In each case, it is critical to decouple the scattering mechanics from the detector response. In this section, we keep this traditional division, although we acknowledge that its meaningfulness begins to dissolve as detector energy thresholds drop below fundamental ionization thresholds.

    \input{Sections/2.1-NR_calibration.tex}

    \input{Sections/2.2-ER_calibration.tex}

    \input{Sections/2.4-Material_properties}

\section{Backgrounds}
\label{sec:backgrounds}

    \input{Sections/3.1-Astrophysical_neutrinos.tex}

    \input{Sections/3.2-Cosmogenic_neutrons.tex}

    \input{Sections/3.3-Radiogenic_neutrons.tex}

    \input{Sections/3.5-Surface_contaminants.tex}

    \input{Sections/3.6-Active_bulk_contaminants.tex}

    \input{Sections/3.7-Near-threshold_phenomena.tex}

\section{Simulations \& Detector-Response Modeling}
\label{sec:simulations}
\textbf{\textit{Lead:} M.\ Szydagis}\\
\textit{Contributors: R.\ Bunker, A.\ Kamaha, S.\ Westerdale, S.\ Sharmapoudel, S.\ Burkhant, \\ A.\ Erlandson, M.\ Kelsey, N.\ Kurinsky}
\\

Calibration-informed simulations of backgrounds and detector-response effects are crucial for future experiments. 
Developing these simulations entails modeling the detector and both internal and external background sources,   
including radiation transport, via Monte Carlo (MC) methods through detector materials into the active medium 
using relevant cross sections from high-energy, nuclear, and AMO physics. 
Contradictory datasets and MC options, insufficient data, and inadequately validated models complicate this task. 
Simulating the atomic-level response of active detector media 
is a separate challenge, including questions of whether elastic or inelastic scattering occurs, and what forms of observable quanta are created from a myriad of processes, such as excitation, ionization, phonons, bubbles, \etc\ 
The efficiencies for generating and detecting these secondaries
have to be simulated, often semi-empirically using combinations of calibrations with first-principle calculations. 
Detectable quanta often have their own transport codes 
followed by detector-specific DAQ simulations. 
Further, final efficiencies can depend on the energy and species of the original primary particle~\cite{SnowmassCF1WP4}. 
There are common issues across different detector technologies, indicative of the need to support a unified effort (see also Ref.~\cite{SnowmassCF1WP4}).

    \input{Sections/4.1-Particle_transport.tex}

    \input{Sections/4.2-Detector_response.tex}

    \section{Facilities and Infrastructure}
\label{sec:facilities}

    \input{Sections/5.1-Radioassasys.tex}

    \input{Sections/5.2-Material_needs.tex}

    \input{Sections/5.3-Material_purity_infrastructure.tex}

    \input{Sections/5.4-Underground_backgrounds.tex}
    
    \input{Sections/5.5-Software_infrastructure.tex}

\bibliographystyle{biblio}
\bibliography{main.bib}

\end{document}

%% file: definitions.tex
\usepackage[utf8]{inputenc}
\usepackage{charter}
\usepackage{fullpage}
\usepackage{graphicx}
\usepackage{lipsum}
\usepackage[version=4]{mhchem}
\usepackage[sort&compress,numbers]{natbib}
\usepackage[colorlinks=true,linkcolor=blue,citecolor=blue,urlcolor=blue]{hyperref}
\usepackage[separate-uncertainty,retain-explicit-plus,per-mode=symbol,binary-units,range-phrase=--,range-units=single]{siunitx}
\usepackage[total={6.5in,9in},top=1in,headsep=0.1in,headheight=1in]{geometry}
\usepackage{enumitem}
\usepackage{amssymb}
\usepackage{subfigure}
\usepackage{comment}
\usepackage{wrapfig}

\newcommand\snowmass{
\begin{center}
  \rule[-0.2in]{\hsize}{0.01in}\\
  \rule{\hsize}{0.01in}\\
  \vskip 0.1in
  Submitted to the Proceedings of the US Community Study\\ 
  on the Future of Particle Physics (Snowmass 2021)\\
  \rule{\hsize}{0.01in}\\
  \rule[+0.2in]{\hsize}{0.01in}\\[-2em]
\end{center}
}

\usepackage[firstpage=true]{background}
\backgroundsetup{contents={\parbox{6.5in}{\snowmass}}, scale=1,placement=top,opacity=1,color=black,position={3.25in,1.2in}}

\usepackage{fancyhdr}
\fancypagestyle{plain}{
  
  \fancyhf{}
  \fancyhead[C]{}
  \fancyfoot[C]{\thepage}
}

\fancypagestyle{empty}{
  \fancyhf{}
  \fancyhead[C]{{\it Snowmass2021 Cosmic Frontier 1: Calibrations and Backgrounds}}
  \fancyfoot[C]{\thepage}
}
\usepackage{authblk}

\usepackage{multirow}
\usepackage[capitalise,noabbrev]{cleveref} 
\newcommand{\dquotes}[1]{``#1''}

\newcommand{\refcite}[1]{Ref.~\cite{#1}}
\newcommand{\refscite}[1]{Refs.~\cite{#1}}
\newcommand{\reffig}[1]{Fig.~\ref{#1}}
\newcommand{\reftab}[1]{Table~\ref{#1}}

\newcommand{\refsec}[1]{Sec.~\ref{#1}}

\newcommand{\eg}{\mbox{\emph{e.g.}}}

\newcommand{\etc}{\mbox{\emph{etc.}}}

\newcommand{\insitu}{\mbox{\emph{in situ}}}
\newcommand{\exsitu}{\mbox{\emph{ex situ}}}
\newcommand{\Insitu}{\mbox{\emph{In situ}}}

\newcommand{\xr}{\mbox{x-ray}}

\newcommand{\gammar}{\mbox{$\gamma$-ray}}
\newcommand{\gammars}{\mbox{$\gamma$-rays}}
\newcommand{\alphap}{\mbox{$\alpha$-particle}}
\newcommand{\alphaps}{\mbox{$\alpha$-particles}}

\newcommand{\cevns}{\mbox{CE$\nu$NS}}
\newcommand{\protonn}{\mbox{$(p,n)$}}
\newcommand{\alphapp}{\mbox{$(\alpha,2p)$}}
\newcommand{\alphan}{\mbox{$(\alpha,n)$}}
\newcommand{\alphang}{\mbox{$(\alpha,n\gamma)$}}

\newcommand{\gamman}{\mbox{$(\gamma,n)$}}
\newcommand{\gammap}{\mbox{$(\gamma,p)$}}
\newcommand{\ngamma}{\mbox{$(n,\gamma)$}}

\newcommand{\nneutron}{\mbox{$(n,n)$}}

\newcommand{\geant}{\mbox{\texttt{Geant4}}}
\newcommand{\fluka}{\mbox{\texttt{FLUKA}}} 
 
\newcommand{\mcnp}{\mbox{\texttt{MCNP}}}
\newcommand{\mcnpx}{\mbox{\texttt{MCNPX}}}
\newcommand{\srim}{\mbox{\texttt{SRIM}}}
\newcommand{\nest}{\mbox{\texttt{NEST}}}
\newcommand{\gfcmp}{\mbox{\texttt{G4CMP}}}
\newcommand{\astar}{\mbox{\texttt{ASTAR}}}
\newcommand{\trim}{\mbox{\texttt{TRIM}}}
\newcommand{\talys}{\mbox{\texttt{TALYS}}}
\newcommand{\empire}{\mbox{\texttt{EMPIRE}}}
\newcommand{\sources}{\mbox{\texttt{SOURCES}}}

\newcommand{\neucbot}{\mbox{\texttt{NeuCBOT}}}
\newcommand{\sagfn}{\mbox{\texttt{SAG4N}}}
\newcommand{\music}{\mbox{\texttt{MUSIC}}}
\newcommand{\musun}{\mbox{\texttt{MUSUN}}}

\newcommand{\ENDF}{\mbox{\texttt{ENDF}}}

\newcommand{\JENDLan}{\mbox{\texttt{JENDL/AN-2005}}}
\newcommand{\livermore}{\texttt{Livermore}}
\newcommand{\penelope}{\texttt{Penelope}}

\DeclareSIUnit\c{\mbox{$c$}}
\DeclareSIUnit\magn{\mbox{$\times$}}
\DeclareSIUnit\min{min}
\DeclareSIUnit\hr{hr}
\DeclareSIUnit\hrs{hrs}
\DeclareSIUnit\week{week}
\DeclareSIUnit\month{mo}
\DeclareSIUnit\months{mos}
\DeclareSIUnit\year{yr}
\DeclareSIUnit\years{years}
\DeclareSIUnit\yr{yr}
\DeclareSIUnit\standard{std}
\DeclareSIUnit\str{sr}
\DeclareSIUnit\ppm{ppm}
\DeclareSIUnit\ppb{ppb}
\DeclareSIUnit\ppt{ppt}
\DeclareSIUnit\pe{PE}
\DeclareSIUnit\spe{SPE}
\DeclareSIUnit\pdm{PDM}
\DeclareSIUnit\ev{events}
\DeclareSIUnit\ct{counts}
\DeclareSIUnit\neutron{\mbox{$n$}}
\DeclareSIUnit\smp{samples}
\DeclareSIUnit\Sample{S}
\DeclareSIUnit\ch{ch}
\DeclareSIUnit\hit{hit}
\DeclareSIUnit\hits{hits}
\DeclareSIUnit\bin{(\mbox{5-PE}~bin)}
\DeclareSIUnit\sgm{\mbox{$\sigma$}}
\DeclareSIUnit\rms{RMS}
\DeclareSIUnit\keVee{\mbox{keV$_{e{\rm e}}$}}
\DeclareSIUnit\keVr{\mbox{keV$_{\rm nr}$}}
\DeclareSIUnit\eVee{\mbox{eV$_{\rm ee}$}}
\DeclareSIUnit\eVr{\mbox{eV$_{\rm nr}$}}
\DeclareSIUnit\ph{photon}
\DeclareSIUnit\el{\mbox{$e^-$}}
\DeclareSIUnit\pm{\mbox{PMT}}
\DeclareSIUnit\pixel{\mbox{pixel}}
\DeclareSIUnit\inch{''}
\DeclareSIUnit\foot{'}
\DeclareSIUnit\bit{bit}
\DeclareSIUnit\sample{samples}
\DeclareSIUnit\barn{barn}
\DeclareSIUnit\bara{bar}
\DeclareSIUnit\bar{bar}
\DeclareSIUnit\barg{barg}
\DeclareSIUnit\mlardepth{\mbox(meter~of~\LAr~depth)}
\DeclareSIUnit\Curie{Ci}
\DeclareSIUnit\psia{psia}
\DeclareSIUnit\psf{psf}
\DeclareSIUnit\pcf{pcf}
\DeclareSIUnit\parsec{pc}
\DeclareSIUnit\cps{cps}
\DeclareSIUnit\mwe{\mbox{m.w.e.}}
\DeclareSIUnit\liveday{\mbox{live-days}}
\DeclareSIUnit\days{\mbox{days}}
\DeclareSIUnit\miles{\mbox{miles}}
\DeclareSIUnit\lumens{\mbox{lm}}
\DeclareSIUnit\degreeC{\mbox{$^{\circ}$C}}
\DeclareSIUnit\degreeF{\mbox{$^{\circ}$F}}
\DeclareSIUnit\electron{\mbox{$e^-$}}
\DeclareSIUnit\Euro{\mbox{\euro}}
\DeclareSIUnit\cph{cph}
\DeclareSIUnit\neq{neq}
\DeclareSIUnit\normal{\mbox{N}}
\DeclareSIUnit\USD{\mbox{\$}}

%% file: author_list.tex
\author[1]{Daniel Baxter\thanks{Co-ordinator, dbaxter9@fnal.gov}}
\author[2]{Raymond Bunker\thanks{Co-ordinator, raymond.bunker@pnnl.gov}}
\author[3]{Sally Shaw\thanks{Co-ordinator, sallyshaw@ucsb.edu}}
\author[4]{Shawn Westerdale\thanks{Co-ordinator, shawest@princeton.edu}}

\author[2]{\\ Isaac~Arnquist}
\author[5,6]{D.S.~Akerib}
\author[7]{Rob~Calkins}
\author[8]{Susana~Cebri\'an}
\author[9]{James~B.~Dent}
\author[2]{Maria~Laura~di~Vacri}
\author[10]{Jim~Dobson} 
\author[11]{Daniel~Egana-Ugrinovic}
\author[12,13]{Andrew~Erlandson}
\author[10]{Chamkaur~Ghag} 
\author[14]{Carter~Hall}
\author[15,16]{Jeter~Hall}
\author[17]{Scott~Haselschwardt}
\author[2]{Eric~Hoppe}
\author[2]{Chris~M.~Jackson}
\author[18]{Yonatan~Kahn}
\author[19]{Alvine~Kamaha}
\author[20]{Mike~Kelsey}
\author[1]{Alexander~Kish}
\author[5,6]{Noah~Kurinsky}
\author[21]{Matthias~Laubenstein} 
\author[5,6]{Eric~H.~Miller}
\author[22]{Eric~Morrison}
\author[22]{Brianna~Mount} 
\author[23]{Jayden~L.~Newstead}
\author[21]{Stefano~Nisi} 
\author[17,24]{Ibles~Olcina}
\author[2]{John~Orrell}
\author[25]{Sergey Pereverzev}
\author[10]{Emily~Perry} 
\author[26]{Andreas~Piepke} 
\author[2]{Sagar~Sharma~Poudel}
\author[27]{Karthik~Ramanathan}
\author[22]{Juergen~Reichenbacher}
\author[28]{Tarek~Saab}
\author[2]{Richard~Saldanha}
\author[4]{Claudio~Savarese}
\author[22]{Richard~Schnee}
\author[15,16]{Silvia~Scorza}
\author[29]{Rajeev~Singh}
\author[1]{Kelly~Stifter}
\author[24]{Burkhant~Suerfu}
\author[30]{Matthew~Szydagis}
\author[1]{Dylan~J.~Temples}
\author[31]{Anthony~Villano}
\author[32]{David~Woodward}
\author[25]{and Jingke~Xu} 

\author[27]{\\ \vspace{0.5em} \textbf{Endorsements:} \\ Taylor Aralis}
\author[33]{Phillip~S.~Barbeau}
\author[32]{M.~Carmen~Carmona-Benitez}
\author[34]{Alvaro E. Chavarria}
\author[35]{Priscilla Cushman}
\author[36]{Kim V. Berghaus}
\author[36]{Rouven~Essig}
\author[5]{Alden~Fan}
\author[37]{Nicolas~Fernandez}
\author[38]{Richard~J.~Gaitskell}
\author[19]{Graciela~B.~Gelmini}
\author[39]{Guillaume~Giroux}
\author[1]{Lauren Hsu}
\author[30]{Cecilia~Levy}
\author[3]{W.~Hugh Lippincott}
\author[40]{Xinran~Liu}
\author[2]{Ben Loer}
\author[17]{Aaron~Manalaysay}
\author[41]{C.~Jeff~Martoff} 
\author[42]{Kirsten~McMichael} 
\author[5,6,43]{Maria Elena Monzani}  
\author[39]{Alexander~StJ~Murphy}
\author[44]{Jong-Chul~Park}
\author[17]{Peter~Sorensen}
\author[17]{Vetri~Velan}
\author[45]{Roc\'{i}o~Vilar~Cortabitarte}
\author[32]{Luis~de~Viveiros}
\author[26]{Juijen~Wang}
\author[46]{Xin~Xiang}
\author[47]{Tien-Tien~Yu}

\affil[1]{Fermi National Accelerator Laboratory (FNAL), Batavia, IL 60510-5011, USA}
\affil[2]{Pacific Northwest National Laboratory, Richland, WA 99352, USA}
\affil[3]{University of California, Santa Barbara, Department of Physics, Santa Barbara, CA 93106-9530, USA}
\affil[4]{Princeton University Department of Physics, Princeton, NJ 08544, USA}
\affil[5]{SLAC National Accelerator Laboratory, Menlo Park, CA 94025-7015, USA}
\affil[6]{Kavli Institute for Particle Astrophysics and Cosmology, Stanford University, Stanford, CA  94305-4085 USA}
\affil[7]{Department of Physics, Southern Methodist University, Dallas, TX 75275, USA}
\affil[8]{Centro de Astropart\'iculas y F\'isica de Altas Energ\'ias (CAPA), Universidad de Zaragoza, 50009 Zaragoza, Spain}
\affil[9]{Department of Physics, Sam Houston State University, Huntsville, TX 77341, USA}
\affil[10]{Department of Physics and Astronomy, University College London, London WC1E 6BT, United Kingdom}
\affil[11]{Perimeter Institute for Theoretical Physics, Waterloo, ON  N2L 2Y5, Canada}
\affil[12]{Department of Physics, Carleton University, Ottawa, Ontario, K1S 5B6, Canada}
\affil[13]{Canadian Nuclear Laboratories Limited, Chalk River, Ontario, K0J 1J0, Canada}
\affil[14]{Department of Physics, University of Maryland, College Park, MD 20742, USA}
\affil[15]{SNOLAB, Creighton Mine \#9, 1039 Regional Road 24, Sudbury, ON P3Y 1N2, Canada}
\affil[16]{Laurentian University, Department of Physics, 935 Ramsey Lake Road, Sudbury, Ontario P3E 2C6, Canada}
\affil[17]{Lawrence Berkeley National Laboratory, 1 Cyclotron Road, Berkeley, CA 94720, USA}
\affil[18]{Department of Physics, University of Illinois at Urbana-Champaign, Urbana, IL 61801, USA}
\affil[19]{University of California, Los Angeles, Department of Physics ${\&}$ Astronomy, Los Angeles, CA 90095-1547, USA}
\affil[20]{Texas A\&M University, Department of Physics and Astronomy, College Station, TX 77843-4242, USA}
\affil[21]{INFN Laboratori Nazionali del Gran Sasso, 67100 Assergi-L'Aquila (AQ), Italy}
\affil[22]{South Dakota School of Mines \& Technology, Rapid City, SD 57701, USA}
\affil[23]{ARC Centre of Excellence for Dark Matter Particle Physics, School of Physics, The University of Melbourne, VIC 3010, Australia}
\affil[24]{University of California, Berkeley, Department of Physics, Berkeley, CA 94720, USA}
\affil[25]{Lawrence Livermore National Laboratory, 7000 East Ave., Livermore, CA 94551, USA}
\affil[26]{University of Alabama, Department of Physics and Astronomy, Tuscaloosa, AL 34587, USA}
\affil[27]{California Institute of Technology, Division of Physics, Mathematics, \& Astronomy, Pasadena, CA 91125, USA}
\affil[28]{Department of Physics, University of Florida, Gainesville, FL 32611, USA}
\affil[29]{Institute  of  Nuclear  Physics  Polish  Academy  of  Sciences,  PL-31-342  Krak\'ow,  Poland}
\affil[30]{The University at Albany, The State University of New York, Department of Physics, Albany, NY 12222-0100, USA}
\affil[31]{Department of Physics, University of Colorado Denver, Denver, Colorado 80217, USA}
\affil[32]{Pennsylvania State University, Department of Physics, University Park, PA 16802-6300, USA}
\affil[33]{Duke University and Triangle Universities Nuclear Laboratory}
\affil[34]{Center for Experimental Nuclear Physics and Astrophysics,
University of Washington, Seattle, Washington, USA}
\affil[35]{School of Physics \& Astronomy, University of Minnesota, Minneapolis, MN 55455, USA}
\affil[36]{C.N.~Yang Institute for Theoretical Physics, Stony Brook University, Stony Brook, NY 11794}
\affil[37]{Department of Physics, University of Illinois at Urbana-Champaign, Urbana, IL 61801, USA}
\affil[38]{Brown University, Department of Physics, Providence, RI 02912-9037, USA}
\affil[39]{Department of Physics, Queen’s University, Kingston, K7L 3N6, Canada}
\affil[40]{University of Edinburgh, SUPA, School of Physics and Astronomy, Edinburgh EH9 3FD, UK}
\affil[41]{Department of Physics, Temple University, Philadelphia, PA 19122, USA}
\affil[42]{Department of Physics, Applied Physics and Astronomy, Rensselaer Polytechnic Institute, Troy, NY 12180, USA}
\affil[43]{Vatican Observatory, Castel Gandolfo, V-00120, Vatican City State}
\affil[44]{Department of Physics, Chungnam National University, Daejeon 34134, Republic of Korea}
\affil[45]{Instituto de F\'{i}sica de Cantabria (IFCA, UC-CSIC), Avenida de Los Castros s/n, 39005 Santander, Spain}
\affil[46]{Brookhaven National Laboratory (BNL), Upton, NY 11973-5000, USA}
\affil[47]{Department of Physics and Institute for Fundamental Science, University of Oregon, Eugene, Oregon 97403}

%% file: Sections/1-Introduction.tex
\section{Introduction}
\label{sec:intro}

This white paper is intended to address the identified community needs related to backgrounds and calibration in future dark matter (DM) direct detection experiments. It should be considered as complementary with the white papers on experimental sensitivity reaching towards the neutrino floor~\cite{SnowmassCF1WP1} and to lower-mass DM~\cite{SnowmassCF1WP2}. It provides parallel discussion of many of the theoretical and modeling needs of the community~\cite{SnowmassCF1WP4} and addresses many of the uncertainties involved in understanding unresolved excess rates in direct detection experiments~\cite{SnowmassCF1WP6}.

\subsection{Executive Summary}
\label{ssec:exec_summary}

Among the needs discussed below, a few themes arise: a need for more nuclear and atomic data, support for and further development of simulation and modeling codes, and investment in underground  infrastructure and advanced detector technologies. These needs become progressively pressing as future direct detection experiments achieve ever-greater sensitivity and search over longer exposures, requiring more precise tools for modeling and mitigating backgrounds and for reconstructing low-energy events.

\begin{itemize}
    \item Calibration measurements are needed of detector responses to electronic and nuclear recoils, for a wider range of targets and at lower energies  (\si{\eV}--\si{\keV} scale and in the ``UV-gap'' of solid-state detectors), including measurements of inelastic and atomic effects (\eg\ the Migdal effect) and coherent excitations, both using established techniques and with new ones (see Secs.~\ref{ssec:NR-calibration}--\ref{ssec:ER-calibration}).
    
    \item Optical and atomic material-property measurements are needed, including atomic de-excitation cascades, electronic energy levels, and energy-loss functions, both for modeling detector response and for simulating transport of low-energy particles. 
    Similarly, improved models and simulations of  detectable-quanta production and propagation are needed  (see Secs.~\ref{ssec:material_properties}, \ref{ssec:astrophysical_neutrinos}, \ref{ssec:near-threshold},  \ref{sssec:particle_transport:geant4}, \ref{sssec:particle_transport:srim}, \ref{ssec:detector_response}).
    
    \item Measurements to decrease neutrino uncertainties are needed, including 
    nuclear-reaction and direct-neutrino measurements that improve simulation-driven flux models (see Secs.~\ref{ssec:astrophysical_neutrinos} and~\ref{sssec:particle_transport:fluka}).
    
    \item Nuclear reaction measurements---\alphan, \ngamma, \nneutron, ($\nu$,$x$), 
    \etc---are needed to improve particle transport codes, background simulations, and material activation calculations, and to validate and improve models/evaluations
    used in these codes; exclusive cross-section measurements are particularly needed to model correlated ejectiles, and codes should provide a full treatment of uncertainties (see Secs.
    \ref{ssec:NR-calibration},
    \ref{ssec:ER-calibration},
    \ref{ssec:astrophysical_neutrinos},
      \ref{ssec:cosmogenic_neutrons}, \ref{ssec:radiogenic_neutrons},
      \ref{sssec:particle_transport:geant4},
    \ref{sssec:particle_transport:fluka}).
    
    \item Increased support for development of simulations (\eg\ \fluka, \geant, \mcnp, \nest, \gfcmp) is needed, both for software codes and models (see Secs.~\ref{ssec:particle_transport}--\ref{ssec:detector_response}).
    
    \item Increased collaboration with the nuclear, atomic, condensed matter, and cosmic-ray physics communities is recommended to improve detector response and background models (see Secs. 
    \ref{ssec:ER-calibration},
    \ref{ssec:cosmogenic_neutrons},
    \ref{ssec:radiogenic_neutrons},
    \ref{sssec:bulk_contaminants:cosmogenic_isotopes}, \ref{ssec:near-threshold}, \ref{sssec:particle_transport:fluka}).
    
    \item \Insitu\ measurements of backgrounds 
    are valuable for validating and improving models in various codes, especially for neutrons and muons
    and material activation measurements, both on surface and underground.
    Likewise, uncertainties on model predictions for these backgrounds need to be quantified by model codes, informed by this validation (see Secs. \ref{ssec:cosmogenic_neutrons},
    \ref{ssec:radiogenic_neutrons},
    \ref{sssec:bulk_contaminants:cosmogenic_isotopes},
    \ref{sssec:other_active_bulk},
    \ref{sssec:particle_transport:fluka}) 
    
    \item Improved material cleaning and screening procedures are needed for dust, radon progeny, cosmogenically activated radioisotopes, and other bulk radioisotopes. New procedures are needed to avoid such contamination, and improved \exsitu\ models of residual background levels are needed (see Secs. \ref{ssec:surface_contaminants}, \ref{sec:active_bulk_fluid}, \ref{sssec:other_active_bulk})
    
    \item Additional R\&D is needed to model, measure, and mitigate near-threshold backgrounds, such as from secondary emission processes (see Sec.~\ref{ssec:near-threshold}).
    
    \item Same-location, multi-method radioassay facilities are needed to simplify measurements of decay chains and reduce systematic errors, alongside greater precision across techniques and increased assay throughput/sensitivity 
    (see Sec. \ref{ssec:radioassays}). Also, development of software infrastructure to track large-scale assay programs across the community is needed (see Sec. \ref{ssec:software_infrastructure}).
    
    \item Finally, investment in underground infrastructure and detector technology (e.g., vetoes) will be needed to mitigate backgrounds such as from radon, dust, cosmogenic activation, and neutrons (see Secs.
    \ref{ssec:radiogenic_neutrons}, 
    \ref{sssec:bulk_contaminants:cosmogenic_isotopes},
    \ref{ssec:material_purity_infrastructure}, \ref{ssec:underground_backgrounds}).
    
\end{itemize}

%% file: Sections/2.1-NR_calibration.tex
\subsection{Nuclear-Recoil Calibration}
\label{ssec:NR-calibration}
\textbf{\textit{Lead:} T.\ Saab}\\
\textit{Contributors: D.\ Baxter, R.\ Calkins, J.\ Dent, A.\ Kish, J.\ Newstead, A.\ Villano}
\\

\begin{figure}
\centering
    \begin{subfigure}{}
        \includegraphics[width=0.7\textwidth]{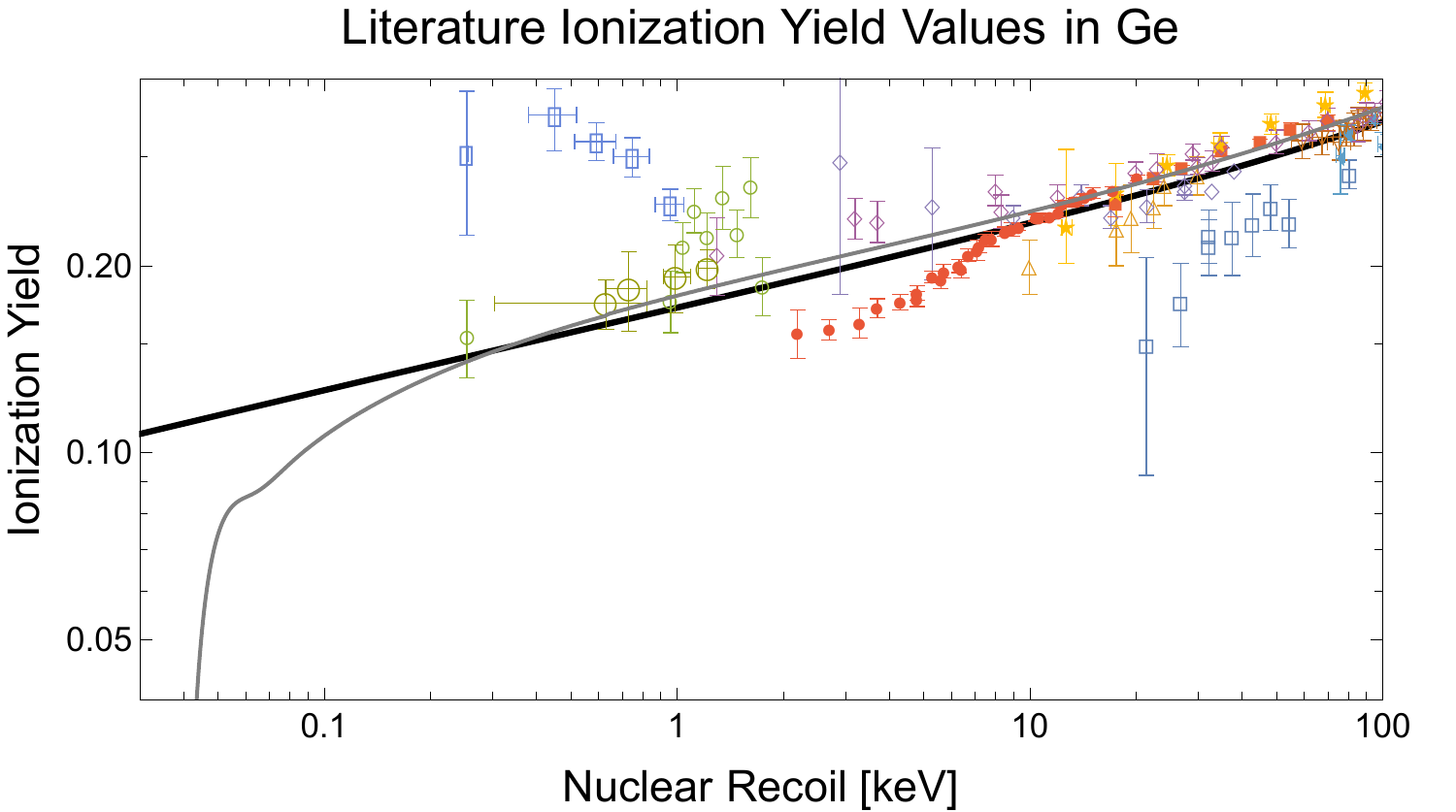}
        \includegraphics[width=0.25\textwidth, trim=0 -10 0 0]{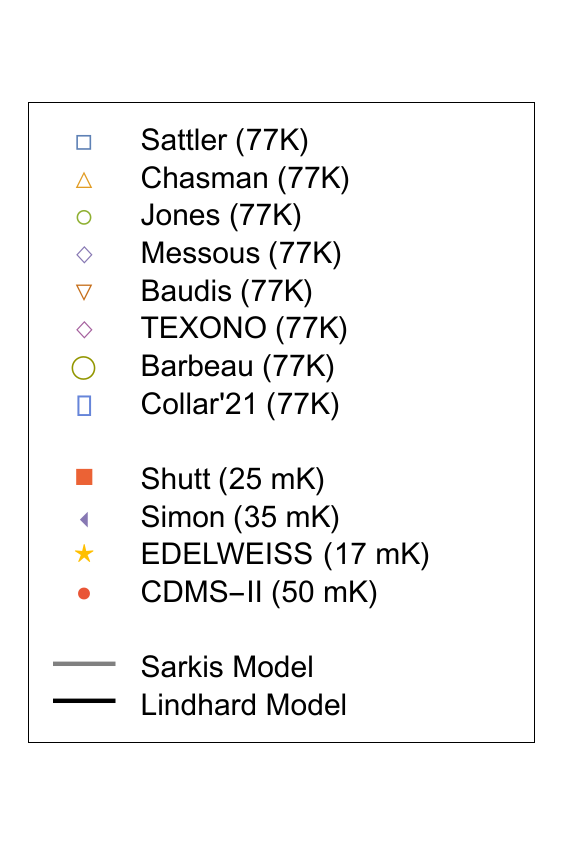}
    \end{subfigure}
\vspace{-0.5em}
\caption{Ionization yield measurements in Ge (right) 
below 100\,keV.  The solid and dashed lines represent the predictions of the Lindhard and Sarkis models~\cite{lindhardIntegralEquationsGoverning1963,Sarkis:2020ds}}
\label{fig:Yield_Ge}
\end{figure}

In many experiments that search for NR interactions of DM, the expected signal is read out via an  
electronic channel such as ionization or scintillation (see, e.g., Refs.~\cite{SuperCDMS:2018gro,DarkSide:2018bpj,XENON:2018voc,AKERIB2020163047,DAMIC:2020cut}). 
Lindhard theory~\cite{lindhardIntegralEquationsGoverning1963} has long been used as the standard model for determining the magnitude (and statistical fluctuation) of the electronic signal as a function of NR energy. 
While this model generally works well for recoil energies above $\mathcal{O}$(10 keV), an increasing number of measurements---in a variety of target materials---have observed deviations from the Lindhard model for lower recoil energies.  
Also, as shown in Fig.~\ref{fig:Yield_Ge}, some measurements 
are inconsistent with each other, which may suggest dependence on detector parameters such as operating temperature or applied electric field. 
Attempts to better model the electronic signal from low-energy NRs include recent theoretical work~\cite{Sarkis:2020ds} to address some deficiencies of the Lindhard model (e.g., lack of atomic binding energy) and 
numerical models developed to estimate the electronic signal in noble-liquid targets~\cite{nest}.

Several calibration techniques, described below, can provide important quantitative information on the detector response to NRs.

\subsubsection{Internal \textit{in situ}}
    One method of obtaining \textit{in-situ} NRs in a target material is making use of the neutron capture reaction; e.g., by exposing a Ge target to a flux of thermal neutrons, the $^{72}\mathrm{Ge}(n, \gamma)^{73}\mathrm{Ge}$ reaction will create Ge nuclei with a mono-energetic recoil energy determined by the emitted $\gamma$. This technique has been used to determine the electronic signal produced by NRs in Ge at the lowest measured recoil energy of 250 eV~\cite{PhysRevA.11.1347,PhysRevD.103.122003}. A variation on this approach is the the Isolated Neutron Capture Technique (INCT). INCT is based around selecting events where the cascading $\gamma$s from the capture event \emph{all escape the immediate region of the NR}, distinguishing it from similar neutron capture methods~\cite{PhysRevA.11.1347,PhysRevD.103.122003}. INCT can provide highly accurate NR calibrations for recoil energies around and below $\sim$100\,eV in light and intermediate mass nuclei. 
    INCT has been demonstrated for Si~\cite{villano2021observation}. The possibility of mid-cascade decay and partial stopping creates a unique recoil spectrum for each  nuclei~\cite{villano2021textttnrcascadesim} with the possibility of a coincidence tag from the exiting $\gamma$.     
    
\subsubsection{External \textit{in situ}}
    NR calibration with an external source offers significant flexibility and a wide variety of sources. The primary challenge with this approach is the continuous neutron energy spectrum from the source and the subsequent recoil spectrum, which makes extracting the NR response as a function of recoil energy challenging and highly dependent on simulation.\looseness=-1
    
    \textbf{Photoneutron sources}: One method of obtaining a mono-energetic neutron source is the photoneutron technique. The approach, which has been used to determine the electronic signal produced by NRs in Si and Ge~\cite{PhysRevD.94.082007,PhysRevD.94.122003,PhysRevD.103.122003,pn2022}, makes use of a Be target combined with a $\gamma$ source to produce nearly mono-energetic neutrons via the two-body reaction $^9$Be($\gamma$,n)~\cite{PhysRevC.94.024613,collar, PhysRevC.85.044605}. The $\gamma$ sources typically used in conjunction with the Be target are $^{88}$Y and $^{124}$Sb. 

    The technique is most effective when it is applied to a detector with a relatively small active volume to minimize the possibility of multiple neutron scatters within the target. It can be applied to larger detectors but careful simulations of the multiple-scatter component are required~\cite{PICO:2019vsc}. 
    The main challenge faced with this technique is the large ratio ($>10^5$) of gammas to neutrons incident on the detector. Significant shielding is required to attenuate the gammas, and Monte Carlo simulations are heavily relied upon to understand the effects of the shielding material on the neutron flux and multiple scattering on the resulting energy depositions.
    
\subsubsection{External \textit{ex situ}}
    \textbf{(p,n) neutron beam}:  
    The ability to perform a calibration at discrete energies is a useful tool that opens the door to precision calibration of the NR response, allowing, e.g., direct measurement of the average ionization produced~\cite{SCENE:2014iyj,Lenardo:2019fcn}, as well an any related statistical fluctuations (e.g.\ Fano Factor), by a NR of a specific energy. This approach makes use of an independent determination of the NR energy of events in a detector based on the scattering kinematics, wherein the neutron beam (of known energy) scatters at a measured angle. 
    
    Mono-energetic neutron beams can be generated at a proton Van de Graaff accelerator via the (p,n) reaction. A variety of targets can be used, in combination with the proton beam energy,  providing significant flexibility in generating a neutron beam of a desired energy. As an example, using the $\mathrm{^7}$Li(p,n)$\mathrm{^7}$Be reaction, a high neutron flux at $\sim$\,500\,keV is readily achievable, with beam energies as low as $\sim$\,50\,keV also  possible~\cite{1949RvMP...21..635H,Beckurts:1964tt}. This relatively low beam energy allows for calibration at sub-100\,eV recoil energies, a regime where it is expected that NRs will cease to produce an electronic signal (i.e., the ionization or scintillation production threshold). Additionally, the collimated nature of the neutron beam enables exploration of potential effects of recoil orientation with respect to crystal planes in crystalline targets. Detecting the scattered neutrons, however, can be a challenge. Selecting recoil energies with high precision requires a relatively small angular acceptance for the scattered neutrons. This means either accepting a small fraction of total interactions in the target or requiring a large array of instrumented neutron detectors.
    
    \textbf{DD and DT neutron generators}: 
     Deuterium-deuterium (DD) and deuterium-tritium (DT) generators are another readily available means for generating a mono-energetic neutron beam. A key advantage is the relative portability of the generator, allowing calibration to be carried out at test facilities or \textit{in situ} at the experiment  facility~\cite{LUX:2016ezw,Verbus:2016xhu,Verbus:2016sgw,Huang:2020ryt}. The 2.5\,MeV neutron energy~\cite{Cranberg:1956kr} from a DD generator can produce a maximum recoil energy in the target material ranging, e.g., from 330\,keV in Si to 75\,keV in Xe. DT generators produce a higher neutron energy of 14.1~MeV. This relatively high endpoint energy allows for NR calibrations to be carried out over a larger range of recoil energies, providing a means to determine the linearity and resolution of the detector response well above measurement threshold. However, this property makes it challenging to calibrate the target near threshold due to the relatively small fraction of small-angle scatters that can produce such events.\looseness=-1
    
\subsubsection{Model calibration of low-energy phenomena (e.g.\ Migdal)}

\begin{wrapfigure}{R}{0.45\textwidth}
\centering
\vspace{-1em}
\includegraphics[width=0.425\textwidth]{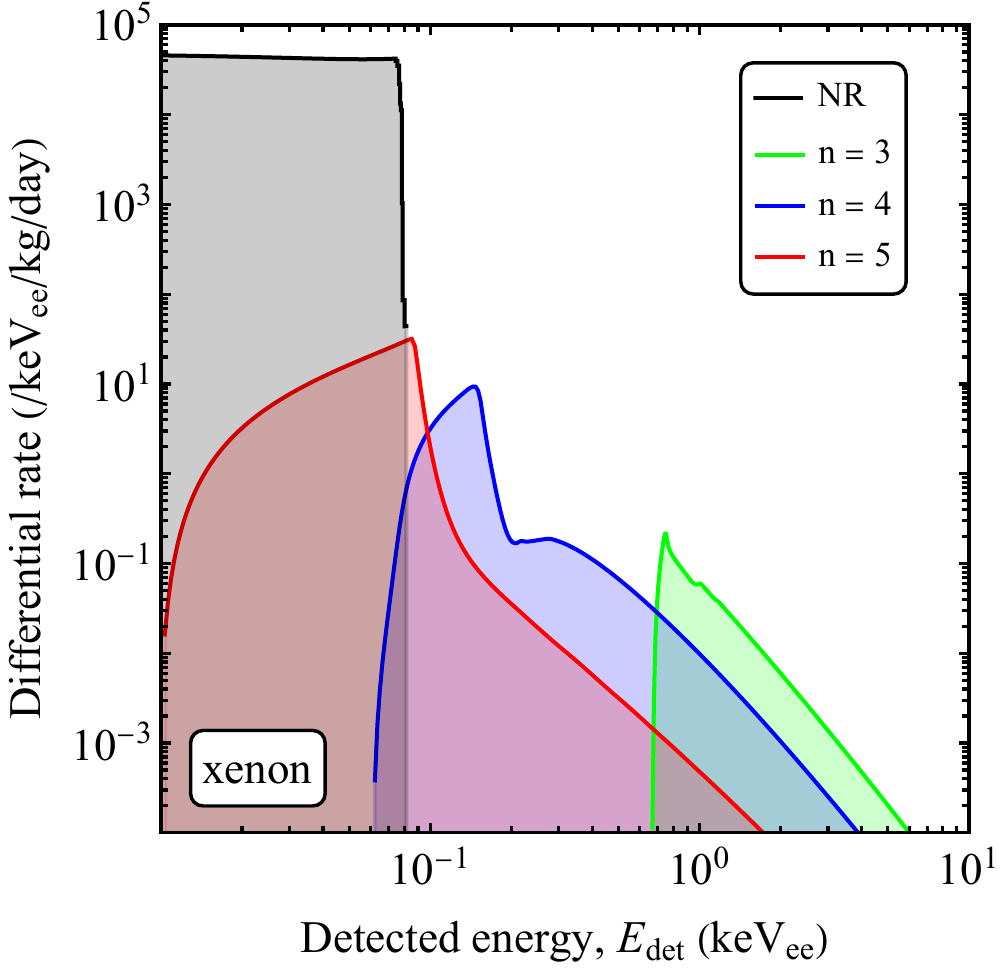}
\vspace{-0.5em}
\caption{Expected scattering rates for elastic NRs and Migdal effect interactions induced by a 17\,keV neutron beam in a LXe target. The Migdal rate is shown separately for the three primary contributing shells. Figure from Ref.\,\cite{bell2021observing}.}
\label{fig:NRCalib_Migdal}
\end{wrapfigure}
    
    Although it is expected that the efficiency for recoiling nuclei to create an electronic signal---via scattering off of neighboring atoms---will decrease with energy and eventually reach zero below some threshold, there are inelastic-scattering channels by which a recoiling nucleus may produce ionization: Bremsstrahlung radiation and the Migdal effect \cite{Kouvaris:2016afs,migdal1939,Migdal:1941}. In the first case, a photon is produced during the DM-nucleus scattering process and can subsequently result in an ER at a nearby location within the target material. In the second case, the low-energy NR perturbs the ambient electron density, resulting in the excitation of electrons above the ground state which may lead to ionization signals. The increased electronic response from these mechanisms can elevate the signal due to very low-energy recoils above the detection threshold, therefore increasing the reach of experiments to lower-mass DM candidates~\cite{Ibe:2017yqa,Dolan:2017xbu,Bell:2019egg}. 
    Theoretical work has shown that the Migdal effect depends on electronic matrix elements related to those governing direct electron scattering, enabling neutron calibration as a potential avenue to validate theoretical models of DM-electron scattering rates~\cite{Baxter:2019pnz,Essig:2019xkx,Liu:2020pat}. The theoretical treatment of these effects generally makes use of the isolated atom approximation, ignoring many-body effects. This may be a reasonable approximation for noble-liquid and gaseous detectors, but is not the case for solid-state detectors where the valence electrons are delocalized. Recent work has taken into consideration the valence states in semiconductors~\cite{Essig:2019xkx} and the effect of the recoiling nucleus on the many-body electron density as encapsulated in the dielectric function \cite{PhysRevLett.127.081805}, demonstrating a potentially large increase in rate compared to the isolated atom treatment.
    
    Demonstrating and quantitatively measuring the Migdal effect in various target materials is a priority; a confirmation effectively increases the low-mass reach of NR DM searches by almost two orders of magnitude in mass. Such a demonstration is challenging because the rate of events due to the Migdal effect is typically 4--5 orders of magnitude lower than the rate of elastic NRs---scaling as $(m_e/m_N)^2$, where $m_N$ is the mass of the scattered nucleus---in the energy range where they overlap. While no unambiguous experimental observation of the Migdal effect from nuclear scattering yet exists, the closely related process of electron shake-off has been observed during nuclear decay~\cite{doi:10.1063/1.1734524,PhysRevC.11.1740,PhysRevLett.108.243201}. One way to boost the Migdal probability is to use MeV neutrons as proposed in Ref.~\cite{Majewski:2021}, which suggests the use of gaseous detectors for which the low density of the gas may allow imaging of both the NR track \textit{and} the electron shake-off. The spectra from Migdal and Bremsstrahlung events extend to higher energies relative to elastic NRs and may be visible above the NR endpoint. Therefore it may be possible to perform an \textit{in-situ} calibration of the Migdal effect inside DM detectors. Feasibility studies indicate an observable rate of Migdal events in the 0.1--10\,keV range with a 17\,keV neutron beam incident on a Xe or Ar target~\cite{bell2021observing} (see Fig.~\ref{fig:NRCalib_Migdal}). At these energies, however, Migdal events may be overwhelmed by background unless precautions are taken with the experimental setup.

    \subsubsection{Phonon production}
    A further, nontraditional type of NR calibration, which will become increasingly important as phonon detector thresholds continue to decrease, is the detector response in the absence of ionization production. While unusual from the perspective of DM physics, neutron scattering measurements of phonon spectra are standard tools of condensed matter physics, and thus calibration techniques may in principle be borrowed and adapted as necessary. Electron energy-loss spectroscopy (EELS) at energies below the gap can also excite phonons; this technique is also useful for calibrating ERs as discussed in Sec.~\ref{ssec:material_properties} below.

%% file: Sections/2.2-ER_calibration.tex
\subsection{Electron-Recoil Calibration} 
\label{ssec:ER-calibration}
\textbf{\textit{Lead:} K.\ Stifter}\\
\textit{Contributors: D.\ Baxter, A.\ Kish, D.\ Temples, K.\ Ramanathan}
\\

A variety of dark-sector models predict observable ER signals~\cite{LOI_szydagis,LOI_ghosh,Alexander:2016aln,Battaglieri:2017aum,SnowmassTF9TheoryLab}, including benchmark models of sub-GeV DM interacting through a kinetically-mixed dark photon. To calibrate these ER interactions, a calibration source with an electromagnetic coupling is required; useful options include $\gamma$s and certain charged particles (e.g., $\beta$s, $\mu$s). Importantly, these options can induce different responses, even for the same incident energy, because, e.g., exciting the target with a 1\,keV photon to initiate an Auger cascade vs.\ injecting a 1\,keV electron may result in different signatures. In any case, accurate knowledge of ER response is required for correct interpretation of experimental results and can be measured in many ways.\looseness=-1

\textit{In-situ} measurements are performed to address a variety of detector-response issues, including linearity and resolution in [$E$, $\vec{x}$, $t$] for ER energy deposits. \textit{In-situ} calibration sources can be external to the active volume---typically point-like encapsulated sources deployed near the active volume---or internally distributed sources, which can include radioactive gases for liquid or gaseous detectors, or activation lines for solid-state detectors.

\textit{Ex-situ} measurements are made to calibrate models that typically encompass two related concepts: (1) the energy deposited for a specific choice of DM particle or background source, and (2) the response of the detector medium to an energy deposition. Similarly, the objectives of these calibrations are twofold: (a) to identify and reject backgrounds (e.g., neutrinos) using detector response signatures, and (b) to inform our expectation of detector response from a variety of signal models.

In the following sections, the variety of calibration techniques used to tackle these questions are summarized, and a more detailed look at the most important questions is given. In all cases, the needs for the coming decade are highlighted. 

\subsubsection{Internal \textit{in situ}}
\label{sssec:intin}

Internal sources, uniquely and crucially, are used for mapping uniformity across the entire active volume of a detector, as well as distinguishing surface from bulk events. Due to the physical nature of various detector targets, there are many different methods (and challenges) associated with calibrating liquid/gas vs.\ solid-state detectors.

For gas- or liquid-phase detectors, radioactive isotopes can be injected directly into the medium via its circulation path such that the isotopes become evenly dispersed within the active volume. Such sources are required to not disturb physics sensitivity. This can be accomplished using \textbf{short-lived isotopes} that decay quickly after a calibration campaign. Examples of commonly used sources are the mono-energetic line from $^{\rm 83m}$Kr~\cite{LUX:2017fwq} and the continuous $\beta$-decay spectra from the $^{220}$Rn~\cite{XENON:2019ykp} chain. 
Looking forward, this technique may become more challenging for kiloton-scale experiments, as the liquid mixing time in the detector may be longer than the lifetime of the isotope. Alternatively, \textbf{longer-lived isotopes} can be injected and subsequently removed. This has been successfully demonstrated with tritium~\cite{LUX:2015amk} and $^{37}$Ar~\cite{XENON:2021fkt}, and could be explored with other sources such as $^{14}$C. The main challenge with this technique is ensuring efficient removal, as any amount of lingering source can poison a DM search.

Rather than injecting radioactive isotopes, sources intrinsically present in experiments can also be used. While this method is also used by liquid/gas-phase detectors~\cite{DEAP:2019pbk,DarkSide:2021bnz}, solid-state detectors rely on it. 
There are several potential intrinsic sources that can be used, including \textbf{fluorescence lines} originating when target materials are exposed to high-energy radiation, 
\textbf{activated isotopes} originating from neutron or cosmic-ray exposure, and \textbf{radioactive decay of contaminants} in a target material.  
A significant drawback of these techniques is the limited selection of energies, which must correspond to known transitions. These sources are also typically reduced as much as possible during detector design, as they also contribute to the background for a DM search. 

As detector thresholds are improved, lower-energy calibrations will be required: sub-keV for liquid/gas detectors, $<$\,100\,eV for current solid-state detectors, and potentially in the meV--eV range for proposed experiments~\cite{SnowmassCF1WP2}. Even in the energy regimes that are currently accessible, an expanded suite of calibration sources would be highly beneficial~\cite{LOI_palladino_gaitskell}. 

\subsubsection{External \textit{in situ}}
\label{sssec:extin}

\textit{In-situ} calibrations can also be performed with external sources. These measurements are often done with dedicated \textbf{radioactive sources}, which have found practical use in calibration because they are inexpensive, can often be used in extended exposures thanks to relatively long half-lives of $\mathcal{O}$(yrs), and can be removed promptly and completely from the detector. These sources are available with a wide range of decay energies from $\mathcal{O}$(1--1000\,keV), including mono-energetic lines and continuum sources. Despite their flexibility, these sources have several drawbacks. There can be environmental hazards posed by high-energy products, as well as subsequent unwanted decays polluting the resulting spectrum. Radioactive sources typically display isotropic emission, which can be an advantage or drawback depending on the application; collimation schemes can be devised if needed. Furthermore, external sources can be strongly attenuated by the support systems that often surround DM detectors, as well as shielded by the outer volume of the target media itself, resulting in distorted energy spectra and varying illumination across the detector. However, fixed sources can have critical uses to measure detector-dependent response and map various localized regions in particularly challenging geometries.

An external calibration source that is common to all experiments is \textbf{cosmic rays}. Requiring km worth of overburden to effectively shield, muons produced in cosmic-ray showers can pass straight through detector media. Advantageously, their angular distribution 
and energies
are known or can be well-measured, such that these quantities can be related back to what is observed in the detector. Further, they can be a source of ``low" energy delta-ray emissions. 
A disadvantage is the obvious limitation of not being able to control the source population of cosmics, which can only be turned off by going underground.

Another, often unwanted, class of external calibrations comes from \textbf{signatures associated with detector construction materials}. This includes fluorescence lines, activation lines, and radioactive decay of material surrounding the detector (used for support or containment) or of surface contamination. 
As these sources represent backgrounds, they will likely be reduced in future detectors and will thus be less useful for calibration.

Another method for external calibration is the use of \textbf{photon sources}, such as lasers or LEDs, which may also include optical fibers, filters, attenuators, collimators, etc.\ placed between the light source and the detector. Advantages include precise targeting of the beam spot, known photon energies, and repeatable measurements. 
Disadvantages in using optical photons include the Poissonian nature of the photon population (and absorption), beam dispersion, aberration, distortion, and the limited choice of photon energies available. Another issue is that they create ``vertical'' electronic transitions---large energy with small momentum transfer---which is different from the kinematics of DM-electron scattering. Also, the absorption coefficients and optical properties of the target can be wavelength dependent~\cite{cohen2012electronic}, which may pose a challenge for new detectors with $\mathcal{O}$(meV) thresholds. Judicious choices of dopants may provide a way forward; e.g., Si:As blocked impurity band detectors can have wideband IR coverage up to 30~$\mu$m wavelengths \cite{woods2018wideband}.

One significant challenge in the landscape of photon sources for solid-state detectors is the \textbf{UV gap}, which refers to the fact that photons in the $\sim$10--100~eV regime typically cannot make it to the detector~\cite{Ramanathan:2020fwm}. For example, 50\,eV photons do not penetrate more than a $\mu$m in most crystals; if the surfaces have other depositions or treatments, the deposit will not occur in the sensitive region of the target. Using lower-energy photons and building up to this energy is a useful workaround but does not directly mimic the instantaneous deposition of the the total energy. 
    
Looking to the future,
detectors for experiments shifting focus to light DM need not be as massive and the external sources described here will be more effective. Still, as thresholds are lowered, it will be necessary to develop methods for lower-energy calibrations. 

\subsubsection{\textit{Ex-situ} model calibration}
    
Despite significant progress in low-energy ER calibrations, there is work to be done in quantifying and modeling the microphysical effects that influence signal production. Across all media, better models of energy deposition, loss, and transfer are needed to enhance predictions of detector response across a wide range of energies and for various classes of ERs ($\beta$-like vs.\
Compton-like) \cite{LOI_westerdale_clark,LOI_mooney,LOI_westerdale}. Thus in the coming decade, as experiments improve in both resolution and threshold, comprehensive experimental programs are needed to answer many questions for ER interactions, such as:
\begin{itemize}

    \item \textbf{Signal yields:} The keys to understanding the conversion between detector observables and deposited energy are absolute light, charge, and phonon yields of a material. For noble liquids, reducing the uncertainty on light and charge yields will be crucial, especially at very low energies~\cite{LOI_westerdale_clark}. These yields also have a dependence on the work function $W$ of the material, and recent measurements ~\cite{PhysRevC.101.065501,Baudis2021a} are in slight tension with the canonical value used by experiments (c.f.\ Ref.~\cite{Dahl:thesis}), which is important to resolve.
    For bubble chambers, yield measurements are still important, but also required is a detailed understanding of the bubble nucleation threshold~\cite{LOI_dahl,PICO:2019rsv}.
    For semiconductors, major targets of understanding are the bandgap $\epsilon_g$, electron-hole pair-creation energy $\epsilon_{\rm eh}$, and Fano factor~\cite{Rodrigues:2020xpt}. The latter two  are statistical quantities, valid by the grace of the central limit theorem at high energies ($E_r \gg \epsilon_g$). For small deposits, the ionization is not Poissonian and model variation of even a few percent can have a tens of percent effect on DM results~\cite{Ramanathan:2020fwm}. Pushing even lower, detector response must also be understood at energies near or below the bandgap (see Sec.~\ref{ssec:material_properties} below). Time-dependent density functional theory (DFT) can provide \textit{ab-initio} modeling of the energy exchange between electrons and phonons following a primary recoil, but measurements in this region are needed to disentangle  simulation systematics from real-world detector effects. 
    
    \item \textbf{Lattice \& atomic structure:} Related to the above item is the question of how atomic structure affects signal yields, which has a particular impact on several key backgrounds. As exposures of liquid-noble DM experiments grow to kiloton-year scales, the impact of the irreducible background of neutrino scattering on sensitivity becomes larger (see Sec.~\ref{ssec:astrophysical_neutrinos}). Historically, a universal ER response has been applied to neutrino-induced ERs, however, at low energies the electronic structure of the target atoms becomes important. A simple, data-driven model of this effect (calibrated using electron-capture sources) predicts only a small impact on overall sensitivity for next-generation DM experiments~\cite{Temples:2021jym}, but the low-side tail of the ionization-to-scintillation ratio has not been measured for this class of events. Internal electron-capture sources can be used to calibrate this effect~\cite{Boulton:2017hub,LUX:2017ojt,Temples:2021jym}. One of the dominant backgrounds for solid-state detectors in the keV range is Compton scattering, which arises from higher-energy photons emitted by background sources scattering off atomic shells and presents a flat background down to the binding energies where there are unique step-like features in the energy spectra~\cite{Ramanathan:2017dfn,Botti:2022lkm}. Modeling and measuring these features and their shapes will be crucial for identifying DM above background. 
    
    \item \textbf{Dopant effects:} The potential benefits of dopants for many target materials has been explored in recent years. In LAr, the addition of a species with lower ionization energy than Ar (such as Xe) increases the scintillation and ionization yield, 
     leading to a reduction in effective threshold~\cite{LOI_westerdale_clark}. In both LAr and LXe, the addition of a species that is kinematically better matched to the mass of the DM model of interest can extend the DM mass range of an experiment, 
    such as the addition of H$_2$ or He to LXe~\cite{LOI_lippincott}.
    Most solid-state crystalline targets are doped (e.g., Si:B or Si:P to create p/n-type materials), and the concentration affects the resisitivity, mobility, etc. of the target, which in turn affects the propagation and lifetimes of excitations and thus directly impacts the detector response (e.g., resolution degradation)~\cite{Fernandez-Moroni:2020abn,DAMIC:2021crr}. In all cases, calibrations are needed to determine the response of the doped medium (to ERs and NRs), as the dopant allows energy partitioning into more degrees of freedom than in an undoped system; some are observable channels, some are not. Calibrations of the yields for ERs in doped systems, as a function of dopant concentration, establish a baseline against which to compare NRs of different species. 
    
    \item \textbf{$\vec{E}$ fields, temperature, and crystal orientation:} There are a variety of external, controllable parameters that affect detector response. For example, it is known that bandgaps change with temperature, but the effect is mostly estimated with numerically fitted models to compensate for the effect~\cite{varshni1967temperature}. Drift experiments use strong electric fields and their effect on things like diffusion and electron recombination lifetimes are usually modeled in an \textit{ad-hoc} manner~\cite{LUXCollaboration2019} and must be better understood as thresholds continue to be lowered~\cite{LOI_westerdale_clark,LOI_mooney}. Finally, recent NR work has demonstrated large differences in quenching factors measured using the same target material at the same energies. One hypothesis is crystal orientation effects, as it is known that electron-hole drift and mobility is different, e.g., in Si $\langle$111$\rangle$ vs.\  $\langle$100$\rangle$~\cite{moffatt2019spatial}. Dedicated campaigns to experimentally determine how these ``state variables'' affect response is needed because solid-state detectors work in many regimes. 
\end{itemize}
    In addition to the techniques discussed in Secs.\ \ref{sssec:intin} and \ref{sssec:extin}, another method for studying these effects is to use \textbf{photon beams}, such as those produced by betatrons or synchrotrons~\cite{ulm2003radiometry}, which can be used to understand ER response at a variety of energies. However, such measurements must occur in high-rate, high-intensity, or otherwise high-background environments, rendering a DM search impossible. Another option, which may be particularly useful at low energies, is \textbf{hot electron injection by way of quantum mechanical tunneling} through tunable potential barriers. This can be a useful tool to study impact ionization or related processes at the eV scale~\cite{chang1985quantum} and may provide a way to inject meV-scale energies into solid-state detectors. Tangentially, recent work has used phonon sensors themselves as a source of energy~\cite{ramanathan2021identifying}. 
    Still, the options at these energies are limited, and with the advent of Quantum Information Science (QIS) influenced detectors (e.g., qubits coupled to substrates) potentially sensitive to meV-scale deposits, a lack of calibration tools in this regime will pose a serious bottleneck for the community. 
    
For the items listed above, those performing dedicated calibrations should work in conjunction with the developers of community simulation tools to ensure results are incorporated in a standard fashion. For further discussion, see Sec.~\ref{sec:simulations} and Ref.~\cite{SnowmassCF1WP4}.

%% file: Sections/2.4-Material_properties.tex
\subsection{Material Property Measurements}
\label{ssec:material_properties}
\textbf{\textit{Lead:} Y.\ Kahn}\\

Dedicated measurements are needed to determine the stopping powers, energy loss functions (ELFs), absorption lengths, and overall cross sections for different interactions in a material for external radiation. Separating DM physics from material physics from detector response during calibration requires excellent knowledge of these material coefficients and can be done independent of DM detector operation, e.g., by studying uninstrumented material response in condensed matter setups.
 
In a generic detector material, the DM-electron scattering rate for interactions where the DM couples to electron density (e.g., in the kinetically-mixed dark photon model which is a common benchmark for sub-GeV direct detection experiments) is determined by the ELF, $\mathcal{W}(\mathbf{q}, \omega)$, which measures the response of the detector material to perturbations of the charge density \cite{Hochberg:2021pkt,Knapen:2021run,Kahn:2021ttr}. Indeed, the ELF takes its name from the fact that it measures the energy lost by an electromagnetic probe (and hence deposited in the material) scattering with momentum transfer $\mathbf{q}$ and energy $\omega$. While there are many models and approximations for the ELF, which is related to the complex dielectric function, a key property of the ELF is that it is experimentally measurable, affording the possibility of a direct calibration with electromagnetic probes of any detector sensitive to DM-electron scattering which automatically accounts for many-body effects. A recent code package \cite{Knapen:2021bwg} allows the input of any measured ELF to compute DM-electron scattering rates. \looseness=-1  

As an example relevant to current experiments, the ELF of Si has been measured using electron beams, in a technique known as electron energy-loss spectroscopy (EELS) \cite{kundmann1988study}, as well as inelastic X-ray scattering \cite{weissker2010dynamic}. The X-ray cross section grows with increasing momentum transfer, and thus the ELF is best measured with X-ray scattering for momentum transfers large compared to the inverse interatomic spacing \cite{weissker2010dynamic}. However, the kinematics of DM, and in particular its slow velocity $v_{\rm DM} \sim 10^{-3}$, are quite distinct from the kinematic regime typically probed in condensed matter experiments. Thus, existing Si ELF data are not sufficient to accurately compute a scattering rate, especially for energy deposits near the gap \cite{Hochberg:2021pkt}. Precise measurements of the ELF in Si (and Ge) in the kinematic regime specific to sub-GeV DM would clearly be of benefit to the community.

Moving to future detectors, a number of groups are investigating materials with sub-eV bandgaps to probe sub-MeV DM~\cite{SnowmassCF1WP2}. In these materials, the momentum transfers may be much smaller than the interatomic spacing, and EELS measurements are the preferred tool to determine the ELF. Further, these compounds often have exotic electronic properties such that many-body effects cannot be neglected, and measurements of the ELF, when available, will furnish an expected signal rate which is robust against theoretical uncertainties. State-of-the-art momentum-resolved EELS (M-EELS) can measure with meV energy resolution and eV momentum resolution \cite{kogar2017signatures}, and is thus ideal for characterizing new  detector materials for DM-electron scattering. Partial information may be obtained with optical spectroscopy, which can measure the ELF at finite energy but zero momentum, but additional theoretical modeling is required to extrapolate to finite momentum.\looseness=-1

%% file: Sections/3.1-Astrophysical_neutrinos.tex
\subsection{Astrophysical Neutrinos}
\label{ssec:astrophysical_neutrinos}
\textbf{\textit{Lead:} S.\ Haselschwardt}\\
\textit{Contributors: I.\ Olcina}
\\

Terrestrial direct detection experiments with sufficient exposure will observe background events originating from interactions of astrophysical neutrinos with the target. 
These events represent an irreducible background, because they cannot be shielded. 
Of particular concern is coherent elastic neutrino-nucleus scattering (\cevns), which can produce recoil energy spectra similar to those expected from some DM signals. 
Systematic uncertainties on these backgrounds, especially due to the neutrinos' flux, will ultimately limit the sensitivity of future direct detection experiments. 
This limitation is often referred to as the \dquotes{neutrino floor}---also known as the \dquotes{neutrino fog}~\cite{OHare:2020lva,OHare:2021utq}---below which the progress of traditional DM searches will be severely hindered~\cite{PhysRevD.89.023524}. 

In this section we discuss the main sources and uncertainties for astrophysical neutrino backgrounds, along with prospects for decreasing their uncertainties. 
Due to their large exposures, we focus on these backgrounds in LXe- and LAr-based experiments.

\subsubsection{Neutrino sources, fluxes, and their uncertainties}

There are three classes of astrophysical neutrinos relevant to direct detection experiments: solar, atmospheric, and diffuse supernova background neutrinos (DSNB). 
Figure~\ref{fig:nu_fluxes} shows the flux-normalized energy spectra of these neutrinos at Earth, and they are discussed in more detail in  \refcite{Vitagliano:2019yzm}. 
Below is an overview of their primary uncertainties:

\begin{figure}[t!]
    \centering
    \includegraphics[width=0.75\textwidth]{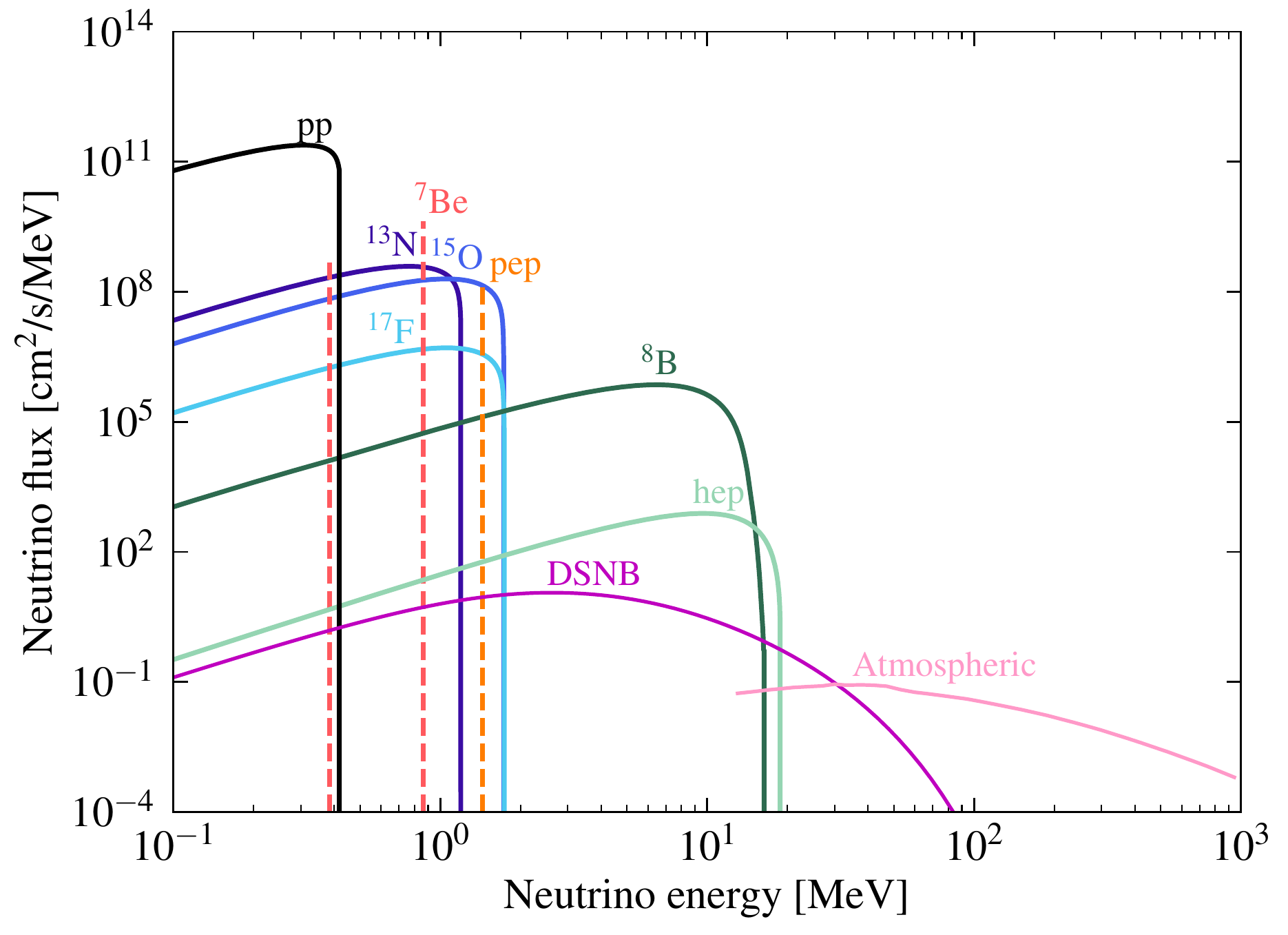}
    \vspace{-0.5em}
    \caption{Dominant neutrino fluxes that constitute a background to direct detection experiments: solar, atmospheric, and DSNB, where each of the individual contributions to the solar neutrino flux are labelled explicitly. Taken from Ref~\cite{Baxter:2021pqo}.}
    \label{fig:nu_fluxes}
\end{figure}

\begin{itemize}
    \item \textbf{Solar neutrinos:} The $pp$ chain has mostly been measured to a few percent uncertainty, and Borexino measured the CNO neutrino flux to \SI{\sim50}{\percent} uncertainty~\cite{BOREXINO:2020aww}. The best theoretical model of the sun is the Standard Solar Model (SSM, see summary in \refcite{Vinyoles:2016djt}) which provides theoretical predictions of the solar neutrino flux with percent-level uncertainties~\cite{1963ApJ...137..344B}. However, inconsistencies between models informed by helioseimology data (low-metallicty model~\cite{Serenelli:2009yc}) and by photospheric measurements of heavier elements (high-metallicty model~\cite{Asplund:2009fu}), give rise to the \dquotes{solar metallicity problem}, which complicates model-driven neutrino flux predictions beyond their stated uncertainties. Resolving this problem and improving CNO flux measurements are therefore important for DM searches below \SI{\sim10}{\GeV}.
    
    \item \textbf{Atmospheric neutrinos:} Atmospheric neutrinos have not been measured below \SI{\sim100}{\MeV}. Fluxes are therefore estimated from simulations, with the current best estimates from 2005 \fluka\ studies~\cite{Battistoni:2005pd}. Although there have been attempts to create new simulations~\cite{PhysRevD.83.123001,Honda:2015fha}, the uncertainties are typically \SIrange{20}{25}{\percent}, mainly from uncertainties on the theoretical interaction cross sections as well as the Earth's geomagnetic field. \fluka\ is further discussed in \refsec{sssec:particle_transport:fluka}. Improving these inputs, measuring low-energy atmospheric neutrinos, and model-building with the atmospheric neutrino community can decrease uncertainties in these backgrounds.
    
    \item \textbf{Diffuse supernova neutrino background (DSNB):} The uncertainty on this flux is dominated by uncertainties in the simulations of stellar core collapse and neutrino oscillation, and is generally considered to be \SI{\sim50}{\percent}~\cite{Strigari:2009bq,Beacom_2010}. Because these backgrounds are generally subdominant to the others, these uncertainties are not a major limiting factor for future direct detection searches.
\end{itemize}

Recently, in an attempt to unify neutrino models and uncertainties among direct detection searches, a large subset of the community conducted a review of measurements and models in \refcite{Baxter:2021pqo}. Recommended values are summarized there (see Table~3 therein).

\subsubsection{Neutrino-electron scattering}
In current and future LXe and LAr detectors, the dominant source of neutrino-induced event rate results from the elastic scattering of solar neutrinos on the bound electrons of the target atoms, as described in \refscite{Marciano:2003eq,Formaggio:2012cpf,Dutta:2019oaj}.

However, there are a few relevant corrections to this calculation. First, the differential cross section is subject to corrections arising from the fact that electrons are not free; they are bound in atomic orbitals of the target atom. One possible solution is to consider a stepping function for atomic shells, but more advanced calculations exist for a Xe target~\cite{Chen:2016eab}. Second, neutrino-flavor transformations  from the solar core to the Earth should be taken into account. The best solution that exists at present is the large mixing angle (LMA)-Mikheyev-Smirnov-Wolfenstein (MSW) model~\cite{Robertson:2012ib,Antonelli:2013ahep}.

\begin{figure}[htb]
\centering
    \begin{subfigure}{}
        \includegraphics[width=0.48\textwidth]{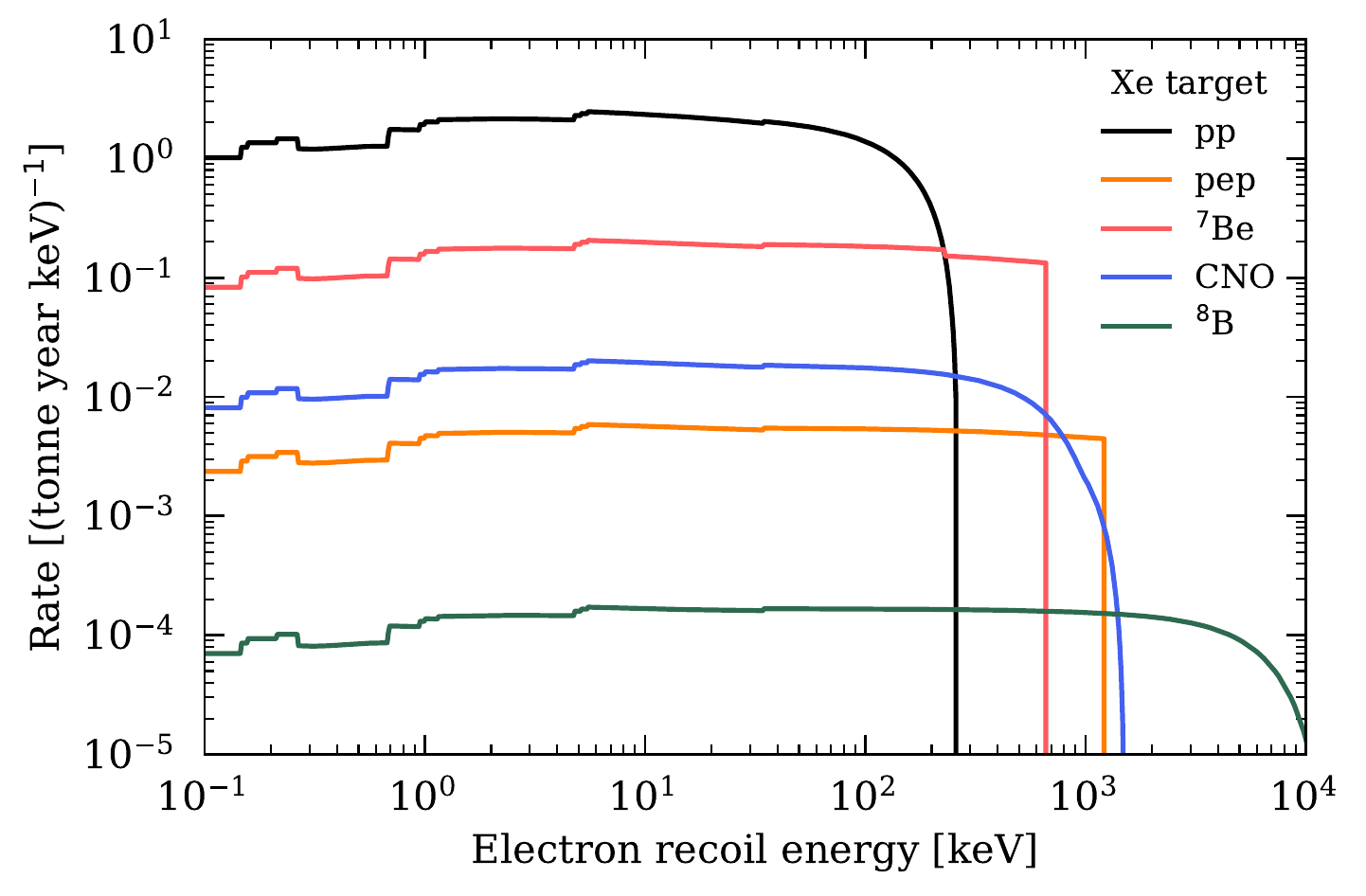}
    \end{subfigure}
    \begin{subfigure}{}
        \includegraphics[width=0.48\textwidth]{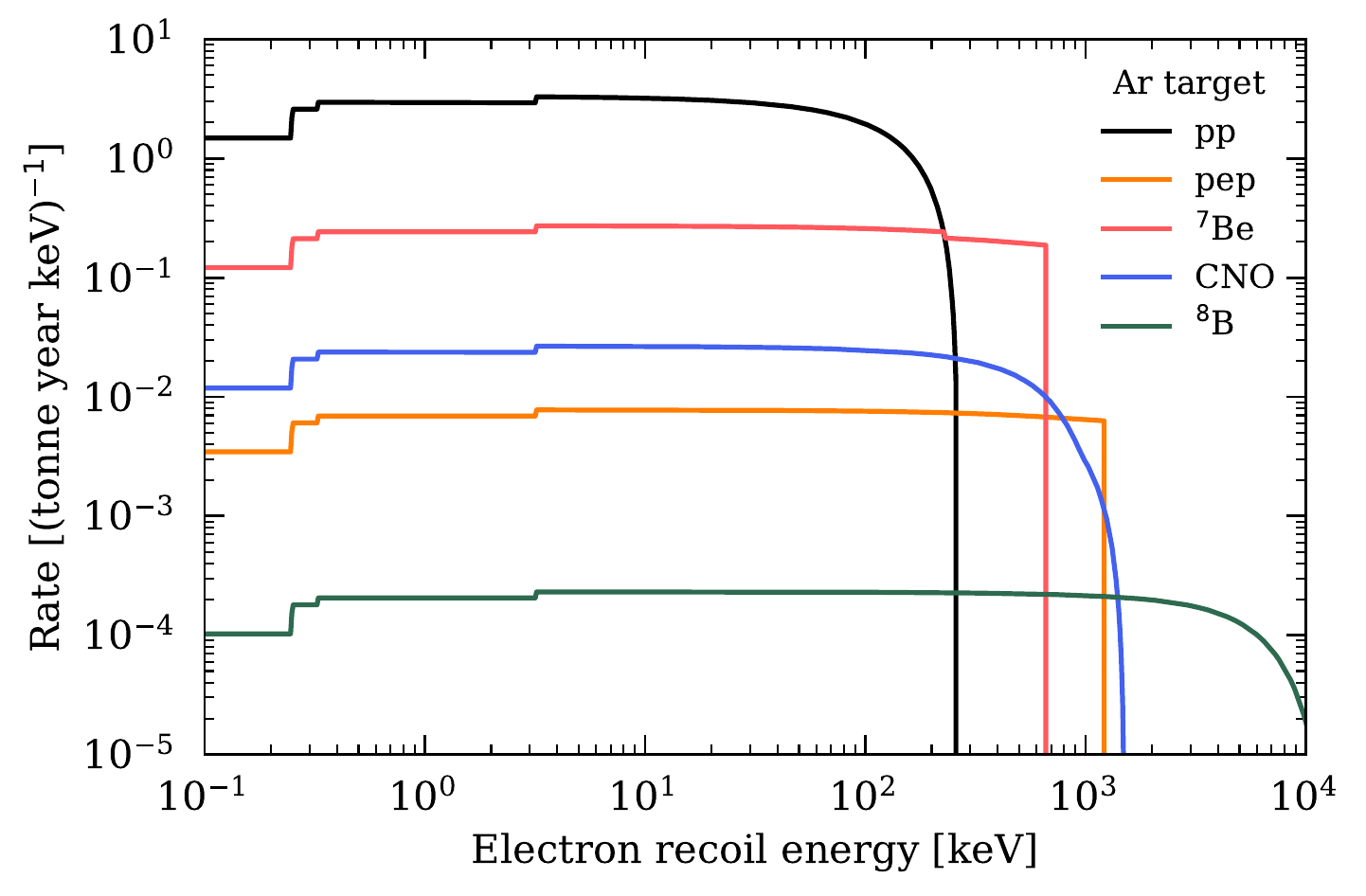}
    \end{subfigure}
    \vspace{-0.5em}
\caption{ER spectra from neutrino-electron scattering on Xe (left) and Ar (right). The relativistic random phase approximation (RRPA) correction on Xe was applied following the prescription from Ref.~\cite{Chen:2016eab}, which stops at $\sim$30 keV. Above this energy, the stepping approximation for atomic binding was used. The stepping approximation is used across the full energy range for Ar.}
\label{fig:nu_er}
\end{figure}

The neutrino-induced electron recoil spectra of a few primary neutrino sources for LXe and LAr targets are shown in \reffig{fig:nu_er}. Solar neutrinos from the \emph{pp} fusion and \ce{^7Be} electron-capture processes contribute the majority of the electron-scattering event rate in the low-energy region of interest (ROI) used for most DM and new physics searches.
A comparison of the expected integrated rates in a typical LXe or LAr TPC is given in \reftab{tab:nu_counts}. Characteristic ROIs for DM searches with each technology were used, but note that these are only illustrative and will vary among experiments.

Discrimination between ERs and NRs in LXe and LAr detectors largely removes these backgrounds from NR-like DM searches. In LXe, this discrimination has been extensively studied over a wide range of detector conditions (\eg~in \refcite{LUX:2020car}); 
however, neutrino scattering on inner shell electrons can slightly weaken it~\cite{Temples:2021jym}.  
In LAr detectors, pulse-shape discrimination can suppress ERs by more than eight orders of magnitude, as demonstrated in DEAP-3600~\cite{thedeapcollaborationPulseshapeDiscriminationLowenergy2021a}. As a result, neutrino-induced ERs are not expected to be a significant background for future LAr-based searches for heavy DM. 
While light DM searches that rely on charge-only channels will not have access to pulse-shape discrimination, ERs from solar neutrinos are still expected to be subdominant to neutrino-nucleus scattering. However, corrections to the electron-scattering cross section in LAr, similar to those developed for Xe in \refcite{Chen:2016eab}, may still be useful for solar neutrino physics through this channel in future detectors. Further work in this area is greatly encouraged.

\begin{table*}[tbh]
  \centering
  \caption{Expected neutrino-induced ER and NR rates on Xe and Ar targets, given in energy regions (in parentheses) typical of those used for DM searches in each target.\vspace{0.5em}}
  \label{tab:nu_counts}
  \begin{tabular}{lll}
    \hline\hline
    \multirow{2}{*}{Target} & \multirow{2}{*}{Integrated ER rate} & \multirow{2}{*}{Integrated NR rate} \\[2mm]
    & $(\text{tonne}^{-1} \, \text{year}^{-1})$ & $(\text{tonne}^{-1} \, \text{year}^{-1})$ \\
    \hline
    Xe & 35.0 (1--15 keV) & 0.05 (6--30 keV)\\
    Ar & 158.7 (10--60 keV) & 0.02 (30--200 keV)\\
    \hline\hline
  \end{tabular}
\end{table*}

\subsubsection{Coherent elastic neutrino-nucleus scattering}

Direct detection experiments are also sensitive to \cevns, which was first observed in 2017 by the COHERENT collaboration~\cite{Akimov:2017ade}. 
A key element of this differential \cevns\ cross section is the form factor, which accounts for the loss of coherence with increasing momentum transfer. 
The most common parameterization is the Helm form factor~\cite{Helm:1956zz}, which describes measurements of interactions with equal coupling to all nucleons.
Because \cevns\ is mediated by the $Z$-boson, which preferrentially couples to neutrons, more precise calculations require individual proton and neutron form factors for each nucleus, and would benefit from the development of more refined models.
Dedicated \cevns\ measurements can provide input needed for neutron form factors, as has been done for \ce{CsI}~\cite{cadedduNeutrinoChargeRadii2018} and \ce{Ar}~\cite{cadedduPhysicsResultsFirst2020}.
Such measurements can also constrain non-standard interactions, which may otherwise make \cevns\ differential cross-section calculations less certain.

\begin{figure}[tbh]
    \centering
    \begin{subfigure}{}
        \includegraphics[width=0.48\textwidth]{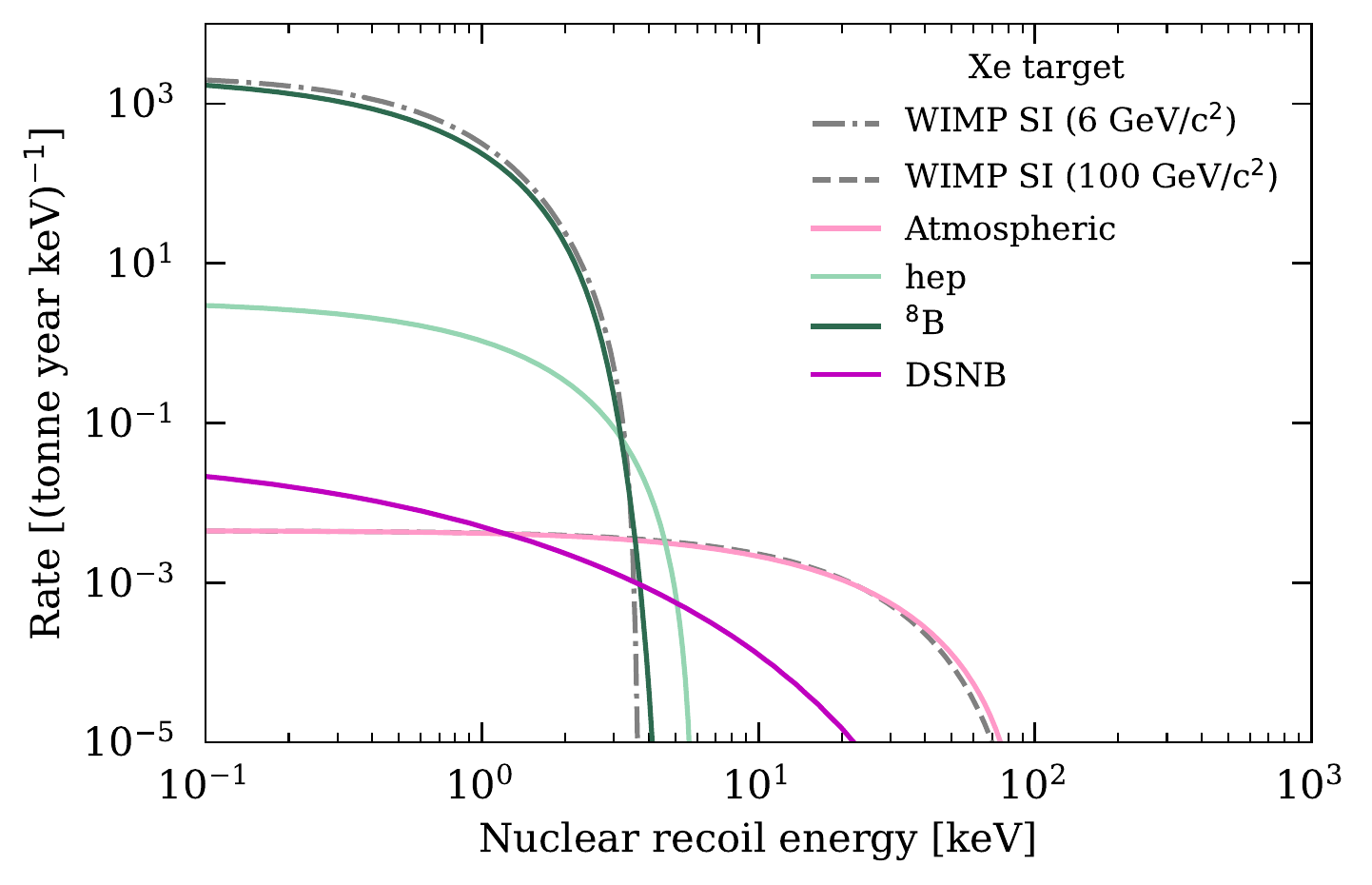}
    \end{subfigure}
    \begin{subfigure}{}
        \includegraphics[width=0.48\textwidth]{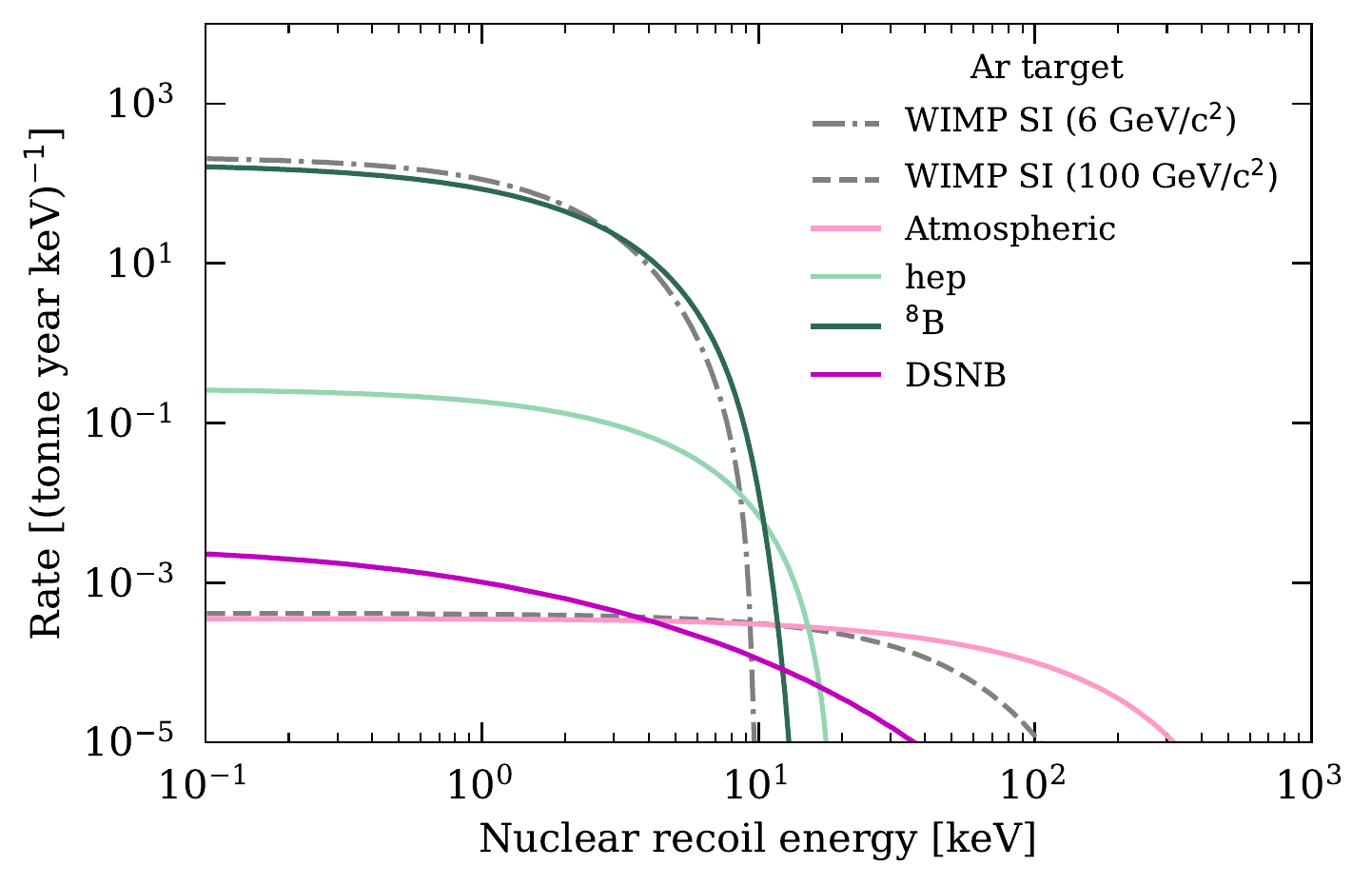}
    \end{subfigure}
    \caption{NR spectra from CE$\nu$NS on Xe (left) and Ar (right). The expected spin-independent (SI) recoil spectra for a 6 and 100 GeV$/c^{2}$ WIMP is also included, showing their significant similarity with the $^{8}$B and atmospheric neutrino spectra, respectively. }
    \label{fig:nu_nr}
\end{figure}

The main sources of \cevns~background in DM searches are the \ce{^8B} and \textit{hep} components of the solar \textit{pp} chain, atmospheric neutrinos, and the DSNB. The predicted NR spectra for each of these sources is shown in Fig.~\ref{fig:nu_nr} for Xe and Ar targets. A striking similarity is observed between the recoil shapes of $^{8}$B and atmospheric neutrinos with those of spin-independently interacting $6$ and \SI{100}{\GeV\per\square\c} WIMPs in LXe, and at slightly different masses in LAr. This has important implications on the ultimate range of WIMP-nucleon cross sections that can be probed by future experiments. This topic is discussed in more detail in \refcite{SnowmassCF1WP1}. It is important to note that a reduction in the systematic uncertainty of the neutrino fluxes would greatly minimize the barriers posed by the neutrino fog (e.g., as shown in Fig.~3 of Ref.~\cite{OHare:2020lva}); we encourage  future effort in this direction.\looseness=-1

\subsubsection{Neutrino capture} 

Electron neutrinos may also be observed through the charged current ``capture'' process on target nuclei resulting in a (possibly excited) product nucleus with increased atomic number and an electron in the final state: $\nu_e + \,^{A}_{Z}X \to \,^{A}_{Z+1}Y^{(*)} + e^{-}$, potentially followed by de-excitation \gammars\ and conversion and Auger electrons.
While this reaction may open the door for interesting solar neutrino spectroscopy, the high energies of the resulting signals and their ER characteristics mean that they are not a significant background for future direct detection experiments.
However, it may be possible to use similar signals to search for Fermionic DM~\cite{Dror:2019onn}, which mimic the neutrino capture process when the DM mass equals the energy of the incoming neutrino. 
For this reason, solar neutrinos may also provide a background to such Fermionic DM searches.
However, as this processes has not yet been observed on argon or xenon and there are significant uncertainties on the reaction cross section and de-excitation cascades, resulting from uncertainties in the structure of the product nuclei (see, \eg, the discussion in \refcite{bhattacharyaWeakinteractionStrengthChargeexchange2009a}).
Additional work modeling this process on argon and xenon would therefore be helpful.

%% file: Sections/3.2-Cosmogenic_neutrons.tex
\subsection{Cosmogenic Neutrons}
\label{ssec:cosmogenic_neutrons}
\textbf{\textit{Lead:} S.\ Poudel}\\
\textit{Contributors: S.\ Westerdale}
\\

Neutrons resulting from cosmic-ray interactions are an important background source, 
including hard neutrons produced in prompt $\mu$ interactions (\eg\ spallation neutrons) and $\beta$-delayed neutrons produced by short-lived isotopes activated by the $\mu$.
Because they may be highly penetrating or produced near the target, these neutrons are difficult to shield.

Traditionally, $\mu$-induced  backgrounds are estimated with detailed Monte Carlo simulations. Commonly used physics codes are \geant~\cite{Geant4_tool_2003}, \fluka~\cite{FLUKA_1,Fluka_2}, and \mcnp~\cite{werner2017mcnp}. 
Simulations require inputs for the $\mu$ flux, which is well-measured at most underground labs, as well as the energy spectrum and angular distribution, which require knowledge of a lab's overburden. 
Accessible overburden models would therefore be helpful; such models would also help with DM calculations probing large cross sections for low-mass candidates, which may be affected by the lab's overburden. 
Energy spectra and angular distributions are often generated by propagating $\mu$s into the lab with software like \music/\musun~\cite{kudryavtsev2003simulations} or by using parameterizations, such as those in \refcite{Mei_2006}. 
Further analyses of the uncertainties inherent to such parameterizations would be helpful.
These simulations often assume a $\mu_+/\mu_-$ ratio \num{\sim1.3}, though more precise measurements, such as those in \refcite{agafonovaMeasurementAtmosphericMuon2010a}, at different labs will improve models.
Many of the calculations rely on nuclear physics that is poorly understood: reaction cross sections often lack data, have large uncertainties, or there are significant conflicts among measurements, and highly uncertain models are often used.  
More measurements to support these models are needed, discussed in \refsec{sssec:particle_transport:fluka}.  
A more thorough understanding of the uncertainty on the data and models is also needed, along with the capability to correctly propagate these uncertainties through calculations; such error bars are often not available for the data and models commonly used. 
Additionally, recent IAEA evaluations rely more heavily on models, for which uncertainties need to be understood. 
\fluka\ simulations also generally require knowledge of exclusive cross sections in order to reproduce correlations among emitted particles, whereas evaluations often provide only inclusive data.
Above \SI{\sim10}{\MeV}, \ENDF\ evaluations become very model-dependent, often with \SI{>30}{\percent} uncertainties; improving these evaluations with data is therefore important.

A faithful simulation requires detailed $\mu$ information (flux, spectrum, angular distribution, multiplicity), proper implementation of composition and geometry of the rock overburden, comprehensive simulation of detector design and geometry, and full transport of the propagating $\mu$s and the resulting secondaries. 
Data on cosmogenic neutrons are scarce, and when data are available, systematic uncertainties on measured quantities are usually large.   
It is important to obtain more measurements to benchmark and validate Monte Carlo codes, including $n$-fold coincidence measurements like those in Refs.~\cite{borexino_2013cosmogenic,kamlandcollaborationProductionRadioactiveIsotopes2010}.  
Additional measurements of underground $\mu$ spectra and angular distributions may also be helpful. 
Often experiments estimate the cosmogenic neutron background using one code or another; analyses using multiple codes would better cover model uncertainties, and comparisons among codes would better benchmark and serve to understand them.\looseness=-1

The $\mu$-induced neutron yield is defined as the number of neutrons produced per fast $\mu$ per unit slant-depth (\eg\  \si{n\per\mu(\gram\per\square\cm)}). 
Empirical scaling laws relating the mean $\mu$ energy $\bar{E}_\mu$ to the neutron yield, such as the $\bar{E}^{0.7}_{\mu}$ relation proposed in \refcite{neutron_yield_1965} or the isotope-dependent power laws described in \refscite{Mei_2006,universal_neutron_yield_2013}, are often used as heuristics, though more detailed models and more thorough understanding of uncertainties are needed to build more precise cosmogenic neutron background models.
Neutron yields are often measured by detecting thermal neutron captures in delayed coincidence with $\mu$ signals, such as those in \refcite{borexino_2013cosmogenic}. 
Where there are a number of measurements of cosmogenic neutron yields deep underground, most are specific to liquid scintillators, systematic uncertainties are often large, and measurements often disagree with simulation models.
Additional measurements will help inform these simulation models, and measurements in other targets like Ar, Xe, Cu, Fe, and Pb will be helpful as well.

Extracting information about cosmogenic neutrons from large-detector data is challenging due to the large $\mu$ signal temporarily blinding the detector to other interactions. 
To precisely measure quantities like neutron flux and yields, advancements in detector technology and event reconstruction may be necessary. 
There are also a few measurements of the cosmogenic neutron energy spectrum underground, such as those in \refscite{LVD_neutron_energy_2009,KARMEN_2003,Soudan_neutron_energy_2014}. 
The cosmogenic neutron energy spectrum is often characterized by a $1/E$ relationship up to a \dquotes{knee}, beyond which it is modeled as $1/E^2$.
Available data are statistically limited and restricted to a small energy range, making it difficult to extract the spectral-shape features. 
More data on cosmogenic neutron energy spectra would therefore be useful. \looseness=-1

Veto detectors can reject $\mu$-induced neutron backgrounds. 
Some of the most challenging cosmogenic neutrons are those from $\mu$ interactions in the rock, where the $\mu$ does not pass through the veto. 
For a given underground site, dedicated efforts to measure the flux and energy spectrum of cosmogenic neutrons emanating from the rock will also help improve  cosmogenic neutron background models. 
Cosmogenic activation on detector materials can also produce short-lived isotopes that emit $\beta$-delayed neutrons, which may be separated from the original $\mu$ by several seconds, making them difficult to tag.
A study in \refcite{empl2014fluka} reports $\beta$-delayed neutron backgrounds from certain isotopes present in liquid scintillator. 
For large detectors, with long run times, it may be important to study the cosmogenic activation in the detector materials to identify if backgrounds from such delayed neutrons have to be accounted for in the background budget.  
Studies like those in \refcite{borexino_2013cosmogenic} will help improve these models in future detectors.

%% file: Sections/3.3-Radiogenic_neutrons.tex
\subsection{Radiogenic Neutrons}
\label{ssec:radiogenic_neutrons}
\textbf{\textit{Lead:} A.\ Villano}
\\

The next decade will require advances in our understanding and
mitigation of radiogenic neutron backgrounds, primarily arising from spontaneous fission, \alphan,
and \gamman\ reactions induced by trace radioactive impurities in detector materials.  Radiogenic
neutrons are typically mitigated through strict radiopurity requirements---informed by modeling the
\alphan\ yield and neutron simulations---and dedicated shielding, including the use of
neutron vetoes~\cite{agnesVetoSystemDarkSide502016}.  These standard avenues need to be
supported fully, and likely other avenues will need to be explored such as dedicated measurements
of key \alphan\ cross sections and dedicated (extremely low) neutron flux measurements through
\ce{^3He}\protonn\ or similar processes.  For many isotopes commonly found in DM
detectors, either with high \alphan\ yields or very high abundance, cross-section
data have \SIrange{50}{100}{\percent} uncertainties at a limited number of energies, and often
these measurements are in conflict with other measurements or have only partial uncertainty
analyses.  For designing future DM detectors and building background models, it is
important to measure these cross sections to the \SIrange{10}{20}{\percent} level.  It
is also important to measure details of correlated \gammar\ emissions.  Backgrounds can be
induced by \gamman\ neutrons, especially near the lower bound of typical DM-search regions of
interest; improved models of this background are needed as well.

\refsec{ssec:radioassays} discusses radiopurity issues, and
\refsec{ssec:particle_transport} discusses uncertainties in neutron propagation.  
In this section we cover models of this process and their associated uncertainties, along with improvements needed, which can be obtained through cross-section measurements, neutron vetoes, and \textit{in-situ} validation of models.   

\subsubsection{Process modeling}
\label{sssec:anmodel}

The \alphan\ neutron flux is typically computed by multiplying a material's \alphan\ yield induced by some $\alpha$-emitting contaminants by the activity of those contaminants, often assuming secular equilibrium throughout some or all of various decay chains.
These calculations require knowledge of (a) the activity of all $\alpha$-emitters present; (b) how the emitted \alphaps\ slow down in the material; (c) the differential \alphan\ cross section; and (d) neutron propagation in the detector (see \ref{ssec:particle_transport}).  
Measurements and improved models in all of these areas are needed. 
A more detailed account of the uncertainties and plans to address them will be provided in an \alphan\ community white paper~\cite{Snowmass-alpha-n}.

Both the ICRU 49 report~\cite{ICRU49}---implemented in NIST's \astar\ tool~\cite{astar} and used in \geant~\cite{Geant4_tool_2003}---and \srim~\cite{srim} offer points
for simulating \alphaps\ slowing down. 
Stopping power calculations are often accurate to within \SI{\sim5}{\percent}, though in composite low-$Z$
materials, the breakdown of Bragg's rule may introduce uncertainties as large as \SI{50}{\percent}
unless chemical binding effects are accounted for~\cite{thwaitesBraggRuleStopping1983}.
Improvements in and/or unification of these treatments will help eliminate aggregate errors down
the modeling chain.

Various codes are available for calculating neutron yields, including
\neucbot~\cite{westerdaleRadiogenicNeutronYield2017},
\sagfn~\cite{mendozaNeutronProductionInduced2020},
\sources~\cite{shoresDataUpdatesSOURCES4A2001,tomaselloCalculationNeutronBackground2008}, and the
USD~calculator~\cite{meiEvaluationInducedNeutrons2009}, which combine various evaluations and model codes, including \JENDLan~\cite{jendl}, \empire~\cite{hermanEMPIRENuclearReaction2007}, and \talys~\cite{koning2008talys}, with stopping power calculations and nuclear decay data.  
Continual maintenance and upgrades to these codes are needed, including improved models driven by new data and additional information about correlated \gammar\ emissions, uncertainties, and information about \alphap\ energy loss prior to capture, among other things.  
Additional comparisons and validation of these codes, such as
in~\refscite{cooleyInputComparisonRadiogenic2018,kudryavtsevNeutronProductionReactions2020} are also valuable.

The codes that are used need the cross sections of many nuclear processes that we collectively
refer to as \alphan\ but could include radiative components like \alphang\ or more than
one neutron like ($\alpha,2n$). Despite the importance of all of these processes it is difficult
to get them all correct in evaluations.  Uncertainties on \alphan\ cross sections are often in the
range of  \SIrange{50}{100}{\percent}, or otherwise ill-defined due to a lack of data or
conflicting measurements; a new set of measurements on relevant materials is therefore needed.
The major challenge for the field is both to continue work to highlight the processes that are the
\emph{most} important and supplement the evaluations with measurements where possible (see
Sec.~\ref{sssec:anmeasure}). 

For rare event searches unambiguously identifying neutrons or \gammars\ from \alphan\ processes would veto the background event. And many collaborations choose to supplement their rare-event detectors with
specifically designed veto detectors. The community, however, must continue to work on correctly
transporting residual neutrons and \gammars\ through geometry to be able to compute the efficacy of
any veto efforts. See \refsec{sssec:anflux} for a brief reflection of the state of the field for
this particle transport. 

Spontaneous fission neutrons are generally well-modeled by the Watt
spectrum~\cite{wattEnergySpectrumNeutrons1952}, and the spontaneous fission yield of \ce{^{238}U}
is well-measured~\cite{browne2015nuclear}.  Due to the higher neutron and correlated \gammar\ multiplicity, these neutrons are more efficiently vetoed in experiments with such a veto~\cite{thesis:Shaw}. However, additional data on the correlation between emitted neutrons and \gammars\ is needed to model this increased efficiency.

\subsubsection{\texorpdfstring{\alphan}{} Measurements}
\label{sssec:anmeasure}

Improving \alphan\ yield calculations requires new measurements.
Current evaluations heavily rely upon models, which may be highly uncertain for reactions of interest in DM experiments, and no direct measurements of \alphan\ yields are available.
Directly measuring \alphan\ reactions on relevant materials and isotopes are therefore needed.
For example, \refcite{Febbraro:2020} measured the \ce{^13C}\alphan\ce{^16O} reaction using Notre Dame's \SI{5}{\mega\volt} terminal voltage accelerated doubly-ionized \ce{He^{++}} atoms (\SI{5.2}{\MeV} and \SI{6.4}{\MeV}) and a \ce{^13C} target, greatly benefiting neutron oscillation experiments and uncovering differences between previous measurements and evaluations.
More measurements like this one are needed~\cite{ReichenbacherIAEA}.

To best use existing facilities for measuring \alphan\ reactions, the community would benefit from developing a prioritized list of reactions where data are needed.
These reactions could then be systematically investigated at several user facilities.
Increased interaction with the nuclear physics community would greatly benefit these efforts.

\subsubsection{Flux measurements}
\label{sssec:anflux}

In next-generation experiments, it has been proposed to validate neutron background models \insitu\ to constrain neutron backgrounds.
For example, it may be possible to use liquid \ce{^3He} scintillation counters---proven technology demonstrated in \refcite{McKinsey:HeScint}---outside a detector to measure the neutron flux, benefiting from the liquid phase's increased density.
Similar validation is also achieved with neutron vetoes. 
These techniques can improve the accuracy of neutron background models and validate radiogenic simulation codes.

\subsubsection{Radiogenic veto possibilities}
\label{sssec:anveto}

The use of a neutron veto to reject radiogenic neutrons was first proposed in \refcite{wrightHighlyEfficientNeutron2011} and developed in
\refscite{westerdalePrototypeNeutronVeto2016,westerdaleQuenchingMeasurementsModeling2017,agnesVetoSystemDarkSide502016}, and has since been adopted by following experiments:  DarkSide-20k~\cite{aalsethDarkSide20k20Tonne2018}, SuperCDMS~\cite{calkinsPrototypingActiveNeutron2015b}, and LZ~\cite{AKERIB2020163047}.  While the basic principle of tagging neutrons by their thermalization and/or capture signals is now well-established, R\&D is needed to address specific needs of experiments.  Also, \alphang\ signals may allow vetoes to achieve higher efficiency by tagging neutrons by their correlated \gammars; this has been studied in simulation for better understood reactions such as \alphan\ on fluorine~\cite{Akerib:2021sims}. However, for many isotopes, current \alphang\ data are insufficient to support such uses.

\subsubsection{Neutron capture backgrounds}
\label{sssec:anncap}

In detectors with sufficiently low thresholds, recoiling nuclei resulting from \ngamma\ reactions
induced by thermal neutrons can produce a potential background.  To mitigate these backgrounds,
emitted \gammars\ must be efficiently tagged in the veto, and the NR response of
thermal neutron-induced backgrounds must be calibrated.  For more information on capturing
neutrons see the discussion of the INCT process in \refsec{ssec:NR-calibration}.

%% file: Sections/3.5-Surface_contaminants.tex
\subsection{Surface Contaminants}
\label{ssec:surface_contaminants}
\textbf{\textit{Lead:} R.\ Schnee}\\
\textit{Contributors: M.L.\ Di Vacri, A.\ Kamaha, E.\ Morrison}
\\

Dust deposition and plate-out of radon progeny are key surface-contamination concerns for detector components. 
For the latter, the long (\SI{22.2}{\year}~\cite{SHAMSUZZOHABASUNIA2014561}) half-life of \ce{^210Pb} causes it to be a background concern long after plate-out.  Betas from \ce{^210Pb} and \ce{^210Bi} on surfaces can cause problematic detector backgrounds~\cite{SCDMSsensitivity,leung2005borexino,xenonEMbackgrounds2011,lux2014backgrounds}.  
Also dangerous are the \ce{^206Pb} recoil nuclei~\cite{SCDMSsensitivity,cresst2012,deap2011surface,deap2015Rn,coupp2012,mount2017lzTDR,Xu2017Ar206Pb} and \alphan\ neutrons~\cite{lux2013nim,lux2015RnBackgrounds,mount2017lzTDR,Breunner2021PTFERnRemoval} that can result from \ce{^210Po} decays. 
Rn progeny can be implanted tens of nm into surfaces  
such that removal via simple surface cleaning is only modestly effective.  
When practical, acid etching can be highly effective at removing \ce{^210Pb} and \ce{^210Bi}, while electropolishing or more careful etching can be effective at removing \ce{^210Po}, which otherwise tends to redeposit~\cite{schnee2013EPRnRemoval,zuzel2015RnRemoval,Guiseppe2018Po210removal,CDMSsidewallEtch2020NIM,Breunner2021PTFERnRemoval,BUNKER2020163870}.  Further work on refining such removal techniques remains important for future DM experiments.

The Jacobi model~\cite{knutson1988modeling,jacobi1972activity}, often with modifications for a cleanroom setting~\cite{leung2005borexino,LRT2017MorrisonPlateout,lz_assay_2020}, can be used to estimate the Rn-progeny plate-out rate~\cite{Stein2018RnPlateout}. 
Covering or enclosing materials can be highly effective at reducing plate-out by greatly reducing the volume of Rn progeny available to plate out. Because \SI{\sim88}{\percent} of \ce{^218Po} atoms are born positively charged in air~\cite{Pagelfopf2005Po218positive}, plate-out onto negatively charged materials is higher than predicted by the Jacobi model. Measurements indicate that materials at the bottom of the triboelectric series (e.g., PTFE~\cite{zou2019triboeletric}) can have a plate-out rate \SIrange{50}{100}{\times} higher than neutral materials~\cite{LRT2017MorrisonPlateout}.  Neutralization of such materials~\cite{lz_assay_2020} effectively reduces this plate-out rate down to levels predicted by the Jacobi model.

\ce{^40K}, \ce{^232Th}, and \ce{^238U} are the dominant contributors to surface radiocontamination from dust fallout. Dust particles generally 
have a chemical composition that reflects the local composition of the soil and which can be affected by anthropogenic activities. The level of \ce{^{40}K} in soil typically ranges from tens to hundreds of ppm, while the \ce{^232Th} and \ce{^238U} content is $\mathcal{O}$(\si{\ppm})~\cite{ncrp1986radiation}, translating to  $\mathcal{O}$(\si{\milli\becquerel\per\kg}) contamination levels that are of particular concern for future DM detectors. In cleanrooms where detector parts are handled, dust  mainly comes from human activities and carry-in particulates.  
To estimate rates and to develop mitigation techniques, research has been dedicated to understand dust composition and fallout, as well as impacts on DM detector backgrounds 
(e.g., Refs.~\cite{di2021direct,parasuraman2012prediction, SNO-STR-91-009, mount2017lzTDR, tiedt2013radioactive, lz_assay_2020}). Further efforts to model and measure dust are ongoing and are important for controlling backgrounds in future experiments.

A key technique for studying dust is to ``witness'' the fallout rate. 
Witness surfaces are used to collect dust in relevant locations and using representative materials. For example, a system of witness plates is in use at SNOLAB to continuously monitor dust fallout.   
Exposed plates are analyzed using X-ray fluorescence (XRF) for mine dust and  
shotcrete via Fe and Ca, respectively (see Ref.\ \cite{boger2000sudbury}).   
Radiocontaminant fallout, in particular from \ce{^232Th} and \ce{^238U}, is inferred from the Ca and Fe,  
based on their relative concentrations in the shotcrete and mine dust~\cite{SNO-STR-2007-003}. Another method is based on the assay of witness surfaces via optical and fluorescence microscopy~\cite{lz_assay_2020}, which is sensitive to particle sizes $\geq$ 0.5\,$\mu$m and provides estimates of the accumulated dust density.  
A recently developed method uses ICP-MS   
to analyze accumulated dust~\cite{di2021direct}.
This method  can be leveraged for more accurate background predictions and mitigation procedures (e.g., quantitative material cleaning);  radionuclide fallout rates  are measured directly. Sensitivities are on the order of \SI{e-3}{\femto\gram\per\day\per\square\cm} for \ce{^232Th} and \ce{^238U} (\SI{\sim e-8}{\micro\becquerel\per\day\per\square\cm}) for a \SI{30}{\day} exposure in a class \numrange{\sim500}{1000} cleanroom. The method can use validated vials as witness surfaces, which can be recapped after exposure and thus allow monitoring of any facility---even those which are not locally equipped with ICP-MS.   
Although extremely sensitive for the determination of \ce{^232Th} and \ce{^238U} fallout, the method does not provide information about the full decay chains and therefore cannot 
verify secular equilibrium in dust. Further R\&D in this area is needed.\looseness=-1

%% file: Sections/3.6-Active_bulk_contaminants.tex
\subsection{Active Bulk Contaminants}
\label{ssec:bulk_contaminants}
\textbf{\textit{Leads:} S.\ Cebrian, E.\ Miller}\\
\textit{Contributors: C.\ Jackson, J.\ Orrell, R.\ Saldanha}
\\
 
The presence of radioactive contaminants in active detector media is a major background challenge. Contaminant levels
typically scale with the active mass and therefore impose strict background controls to limit their impact on DM sensitivity. Special production and purification methods have been considered or are in development  to suppress these isotopes (see Sec.\ref{ssec:material_purity_infrastructure}). There also remains uncertainty in the low-energy spectra for some isotopes, in particular for beta emitters, highlighting the need for further measurements to enable accurate modeling of residual backgrounds in DM-search analyses~\cite{mougeotConsistentCalculationScreening2014,mougeotReliabilityUsualAssumptions2015}.

\subsubsection{Cosmogenic isotopes}\label{sssec:bulk_contaminants:cosmogenic_isotopes}

Long-lived radioisotopes produced via exposure to cosmic rays can be problematic. The spallation of nuclei by high-energy nucleons is the dominant process for this cosmogenic activation, but other reactions are also relevant.  
While there are some direct measurements of production rates, in many cases they must be evaluated from the flux of cosmogenic particles, $\phi$, and the isotope-production cross section, $\sigma$:
\begin{equation}
R=N_t\int\sigma(E)\phi(E)dE ,
\label{eq:cosmo_prod_rate}
\end{equation}
\noindent with $N_t$ the number of target nuclei and $E$ the particle energy. Activation at the Earth's surface is typically dominated by the flux of cosmic-ray neutrons, which 
can be parameterized from MeV to 10~GeV (as in Refs.~\cite{ziegler,gordon}). Other tools include EXPACS  
\cite{expacs}, which calculates particle fluxes for different positions and times in the Earth's atmosphere, and the EXFOR database \cite{exfor} which compiles production cross-section measurements (though neutron-induced cross sections at $\gtrsim$ 100 MeV are relatively rare). Estimates of cross sections can also be obtained from semiempirical formulae \cite{tsao1,tsao2,tsao3} implemented in codes like COSMO \cite{cosmo} and ACTIVIA \cite{activia}, or from hadronic simulations using different codes \cite{boudard2013new, koning2008talys}; some libraries offer production cross sections computed for different targets, products, and projectiles \cite{tendl,jendl,head2009}. Studies of cosmogenic activation for different materials relevant in rare-event searches are reviewed in Refs.~\cite{cebriancosmogenic,kudrylrt2017,universe6100162}. 

Experiments using crystalline detectors have identified a number of  potentially problematic activation products. Several isotopes activated in \textbf{germanium} have half-lives $>$20\,days (see Refs.~\cite{WOS:A1992JZ40300045,Aalseth:11prl2,AGNESE20191,WOS:000401213600007}). While many decay via electron capture---producing spectral peaks that are generally distinguishable from a potential DM signal---several are $\beta$-decay isotopes resulting in spectra that are more difficult to distinguish (e.g., $^{3}$H, $^{63}$Ni, $^{60}$Co, $^{45}$Ca, and $^{22}$Na)~\cite{WOS:000442417200001,AGNESE20191}. In \textbf{silicon}, there are only a few cosmogenically activated radioisotopes, with the most dangerous being $^{3}$H and $^{22}$Na; activation rates have been measured on a neutron beam \cite{saldanha2020cosmogenic} and they have been identified in Si CCDs at rates consistent with the cosmic-ray exposure  \cite{aguilar2021characterization}.  Measurements with \textbf{NaI(Tl)} detectors have revealed several activation products (see Refs.~\cite{naijcap,naireview,naicosine,naidama}). Similar to Si, the production rates of $^{3}$H and $^{22}$Na are important for DM searches. The presence of $^{22}$Na has been evaluated~\cite{anaisbkg2019,naicosine} and a $^3$H rate compatible with observed yields has been computed~\cite{tritiumpaper}; research is ongoing to directly measure the $^{3}$H production rate in NaI. 
For \textbf{CaWO$_{4}$}, $\gamma$ lines from several cosmogenic isotopes were identified \cite{cresstbkg} and a study is underway comparing Geant4 simulations (using ACTIVIA) with data taken with several crystals~\cite{cressttaup}. $^{3}$H, $^{179}$Ta, and $^{181}$W production is the dominant concern for DM searches using CaWO$_{4}$.

As DM detectors grow more sensitive and background requirements get ever more stringent, consideration of cosmogenic activation of detector materials becomes increasingly important. Here we outline areas where careful consideration and additional work is needed to better understand and limit this dangerous source of background.
\begin{itemize}
    \item \textit{Measurement of cosmogenic activation rates}:
    There are uncertainties in calculations of activation yields. Increasing the availability of direct measurements performed under controlled conditions is needed to validate models of production cross sections.
    \item \textit{Verification of exposure models}: 
    Activation rates depend directly on cosmic-ray particle fluxes (see Eq.~\ref{eq:cosmo_prod_rate}). While several models of the flux variation with latitude, longitude, and altitude exist, to our knowledge there has not yet been a careful verification using well-controlled exposures and known cross sections. 
    \item \textit{Use of shielding and underground storage}: 
    Future DM experiments will need to track exposure of detector materials from fabrication to deployment  (as in Refs.~\cite{ABGRALL201552, WOS:000401213600007, WOS:000442417200001}). To reduce activation, demonstrated mitigation methods should also be considered, including shielded transport~\cite{BARABANOV2006115} and shallow underground storage (with studies to evaluate overburden required to reach target background levels). 
    \item \textit{Suppression and removal of activation products}: It is commonly assumed that crystal growth drives out other elements and resets the exposure clock (e.g., for $^3$H), but  material-dependent removal efficiencies have not been quantified. Infrastructure for underground crystal growth and detector fabrication may be required for strong cosmogenic-isotope suppression~\cite{osti_1424835,978-92-79-08276-4,osti_1463301}. It may also be feasible to remove activation products via post-processing, such as baking at elevated temperatures~\cite{sopori2001silicon, ichimiya1968solubility, qaim1978triton}.  Some of these methods may require substantial new infrastructure or R\&D.
\end{itemize}

\subsubsection{Radioactive isotopes in fluid detectors} \label{sec:active_bulk_fluid}
Bulk contaminants in fluid detectors are generally limited to those which dissolve into the fluid and cannot be chemically purified (e.g., via continuous circulation); typically noble elements such as Ar, Kr, Xe, and Rn.  
It is noteworthy that bubble chambers are capable of a high degree of rejection of bulk backgrounds~\cite{ PhysRevD.100.082006,Archambault_2011}. Nevertheless, impurities in the bulk can cause bubble nucleation sites~\cite{ PhysRevLett.118.251301}. There is a general need to control bulk contaminants in liquid detectors.

An equilibrium concentration of Rn can be sustained by decays of progenitor isotopes in detector materials via Rn emanation.
Radon is produced with $\mathcal{O}(100)$~keV of kinetic energy, which is sufficient to travel a few $\mu$m and potentially be ejected from a material into the detector medium. Radon can also diffuse out of a material and into a liquid medium, with a characteristic diffusion length of $L = \sqrt{D(T)/\lambda}$; $D(T)$ is the temperature-dependent diffusion constant (cm$^2$/sec) and $\lambda$ is the Rn mean life. 
 This length ranges from $\mathcal{O}(1)$ mm for $^{222}$Rn in room-temperature plastics (e.g., 2.2\,mm in HDPE\cite{Mamedov_2011}) to approximately zero for most metals.   
In experiments with circulation systems, components such as pumps and getters may contribute significant Rn emanation far away from the instrumented detector, possibly with a relatively small contribution from material surfaces in direct contact with the active medium (see, e.g., Ref.~\cite{XENON:2020fbs}). So-called ``naked'' $\beta$ decays from Rn progeny such as $^{214}$Pb are a key background concern for liquid-based DM detectors.\looseness=-1

Krypton has no naturally-occurring radioisotopes; however, \ce{^85Kr} is present due to nuclear weapons testing and fuel reprocessing \cite{collon2004radionuclides,AHLSWEDE201334}.  
\ce{^85Kr} is background concern because it is a long-lived (T$_{1/2}$=\SI{10.8}{\year}) $\beta$ emitter.   
The abundance varies but is typically within \SI{20}{\percent} of \num{2e-11} in atmospheric samples \cite{AHLSWEDE201334, du2003krypton}. 
Due to the challenge of measuring this low level of \ce{^85Kr}, \ce{^{nat}Kr} (\SI{1}{\ppm} abundance in air) is used as a tracer from which the \ce{^85Kr} level is extrapolated.  
Research grade Xe typically has $\mathcal{O}($\numrange{1}{100}$)$\,\si{ppb} \ce{^{nat}Kr}/\ce{Xe}, while LXe-based DM experiments require \SI{<1}{\ppt}. 

Argon isotopes of concern are \ce{^37Ar}, \ce{^39Ar}, and \ce{^42Ar}. \ce{^37Ar} is produced by interactions of cosmogenic neutrons on atmospheric Ar (AAr) and has a short half-life (35 days); so, activity in a DM detector is typically not a concern after a few months of deployment underground~\cite{agnes2016results}. \ce{^37Ar} has also been used as a calibration source \cite{sangiorgio2013first, akimov2014experimental}. The sea-level cosmogenic production rate in Ar has been measured using neutron beam irradiation \cite{saldanha2019cosmogenic}, and the production rate in Xe has been estimated in \refcite{aalbers2022cosmogenic}. \ce{^39Ar} is a pure $\beta$ emitter that is relatively long-lived (\SI{268}{\year}) and is a dominant background source for Ar-based DM detectors. \ce{^39Ar} is primarily produced by the interactions of cosmogenic neutrons on AAr and is present at the level of \SI{\sim 1}{\becquerel\per\kg}~\cite{loosli1968, benetti2007}. Next-generation DM detectors propose to use Ar extracted from deep underground (UAr). While \ce{^39Ar} production mechanisms also exist underground~\cite{mei2010prediction}, one source of UAr was demonstrated to have an activity of \SI{7.3e-4}{\becquerel\per\kg}~\cite{agnes2016results}. 
New facilities are under development to extract and purify several tonnes of UAr from this source~\cite{aria}. Cosmogenic activation during transportation and storage is also a concern~\cite{saldanha2019cosmogenic}, and new methods to screen UAr are under development~\cite{dart}. \ce{^42Ar}, predominantly produced by \ce{^40Ar}\alphapp\ interactions in the atmosphere, is present in AAr at the level of \SIrange{40}{168}{\micro\becquerel\per\kg}~\cite{agostini2014background, barabash201639, ajaj2019electromagnetic}. \ce{^42Ar} has a similar spectral shape to \ce{^39Ar}, but it is a sub-dominant background as a result of its much lower activity. However, the short-lived \ce{^42K} decay product is key background concern for rare-event searches at high energies (\eg, neutrinoless double-$\beta$ decay and solar neutrinos).

\subsubsection{Other long-lived isotopes in active detector bulk}\label{sssec:other_active_bulk}
Experimental efforts often select active detector media that are already highly purified. The above two subsections cover cosmogenic isotopes generated in a material \emph{after} commercial production and isotopes that are entrained into (fluid) detector media. Beyond these there are a few additional long-lived radioisotopes which have been identified in the active bulk of detectors; we cover these here.

Highly purified Si contains trace levels of $^{32}$Si, which is entrained in raw Si ore and then retained through the refinement process~\cite{ORRELL20189}. 
Reported \ce{^32Si} levels vary, \eg, from $80^{+110}_{-65}$\si[per-mode=power]{\per\kg\per\day}~\cite{Aguilar-Arevalo2015} to \SI{140\pm30}{\micro\becquerel\per\kg}~\cite{DamicCollaboration2021}. This variability is likely linked to the original source of Si ore. \ce{^32Si} is produced in the atmosphere via spallation on \ce{^40Ar}, and it is believed precipitation levels, local soil/strata effects on \ce{^32Si} entrainment, and specific ore collection locations (\eg, dry hills \textit{vs.}\ river beds) all impact the resulting \ce{^32Si} levels~\cite{ORRELL20189}.

The \ce{^32Si} measurements in \refscite{Aguilar-Arevalo2015,DamicCollaboration2021} use a time-correlation, location-coincidence method in CCD detectors, which provides powerful background rejection in DM searches. However, other Si-based DM experiments anticipate the presence of \ce{^32Si} is a potentially dominant background~\cite{SCDMSsensitivity}. The variability of measured \ce{^32Si} levels suggests that a screening program may be able to identify a source of Si ore with suppressed levels. Alternatively, the Avogadro project~\cite{Becker1995522,Becker201049,Fujii2016A19}---focused on producing a kilogram standard---demonstrated the ability to produce isotopically enriched \ce{Si}. The enrichment process is likely highly effective in reducing the \ce{^32Si} concentration; a recent report suggests the Avogadro material is likely already sufficiently pure to reduce the \ce{^32Si} background below the rate expected from solar neutrinos~\cite{ORRELL20189}. Thus, \ce{^32Si} can mitigated, though the reduction techniques will require investment in screening methodologies and/or isotopic enrichment programs. Curiously, there may be crossover utility for \ce{Si} isotopic enrichment in the field of quantum computing to eliminate \ce{^29Si}, a spin-1/2 nucleus~\cite{Dwyer_2014,WOS:000500497100001}, potentially providing an additional motivation for  infrastructure investment.

\ce{^40K} and \ce{^210Pb} in NaI(Tl) detectors, at the present radiopurity levels achieved, are key background concerns \cite{anaisbkg2019,cosinebkg}. Research is ongoing to reduce the activity of these isotopes \cite{cosine200,PhysRevResearch.2.013223}, with promising results based on re-crystallization and zone-refining procedures \cite{PhysRevApplied.16.014060}, and is expected to continue into the future.  Sensitive techniques capable of assessing the low values achieved---\SI{0.1}{\milli\becquerel\per\kg} or below---must be envisaged too.

%% file: Sections/3.7-Near-threshold_phenomena.tex
\subsection{Near-Threshold Phenomena}
\label{ssec:near-threshold}

\subsubsection{Spurious electrons in noble-liquid TPCs}
\label{sssec:spurious_electrons}
\textbf{\textit{Lead:} J.\ Xu}
\\

Noble-liquid TPCs have world-leading sensitivity to low-mass DM through ionization-only searches. Ionization can be collected with \SI{\sim100}{\percent} efficiency, and the electroluminescence gain enables detection of single electrons.  
Experiments have leveraged this efficiency to lower thresholds and search for sub-\si{\GeV} DM interactions (see, \eg, \refcite{XENON:2021myl}). 

A major challenge for improving sensitivity is the few-electron background, which typically manifests as a quickly rising event rate below \numrange{5}{6} electrons and a relatively large single-electron peak. Without accompanying scintillation signals, the nature of these events is difficult to determine. Leading hypotheses include capture and  release of electrons by impurities, delayed emission of surface-trapped electrons at the liquid-gas interface, and electron emission from metal electrodes.  
It is well-known that electronegative impurities in \ce{Ar} and \ce{Xe} can capture drift electrons. Recent observations suggest that some electrons may be released up to a few seconds later and thus contribute to the spurious-electron background. This component increases with the impurity concentration and the electron drift time in the liquid; resulting electrons usually correlate in position and intensity with the progenitor ionization event. 
However, a one-to-one mapping between the electron capture rate and the electron background rate has not been demonstrated. 

Surface-trapped electrons and metal-surface emissions can produce multiple-electron events, which are more dangerous for DM searches. A strong field is needed in a dual-phase TPC to accelerate electrons past the energy barrier at the liquid-gas interface; otherwise, electrons can be trapped under the liquid surface. It is speculated that they may be later released, but exact mechanisms are not yet fully understood. In Ar detectors the required extraction field is relatively low and this background is suppressed. Also, use of high fields in \ce{Xe} TPCs appears to mitigate this background. Other potential mitigations include fast high-voltage switching to redirect trapped electrons and emission stimulation with infrared light. 
As for electron emission from metal surfaces, the field strength at electrode surfaces in a typical TPC is far below the threshold to initiate significant emission; however, intense emission of single- and multiple-electrons has been observed. Although the exact mechanism for this emission is not well-understood, surface treatments (\eg, passivation) may be very effective in reducing emission. Nevertheless, this emission mechanism is expected to be a key background challenge in future DM experiments.

\subsubsection{Secondary emission processes}
\label{sssec:secondary_emmission}
\textbf{\textit{Lead:} D.\ Egana-Ugrinovic}
\\

Energetic particles passing through detector materials lead to secondary emission of low-energy quanta. These secondaries are a near-threshold background concern  in low-mass DM experiments. While some of these events can be effectively vetoed, secondary emission can also arise from interactions in uninstrumented parts of the detector.

Secondary photons are emitted from charge tracks in dielectrics via Cherenkov, transition, and diffraction radiation. Amongst these processes, Cherenkov radiation is especially relevant as a background concern, typically producing photons with \si{\eV}-scale energies; however, estimates must be done on a case-by-case basis~\cite{Du:2020ldo}. Note that Bremsstrahlung radiation is suppressed with respect to Cherenkov emission by powers of the fine structure constant, but it can dominate for photon energies above a few \si{\eV}. 

Secondary emission also occurs \textit{indirectly} via excited electronic states in materials, which can emit light or heat in the process of relaxing back to the ground state. For photon emission, this process goes under the names of scintillation, luminescence, or phosphorescence (or afterglow); if the relaxation involves the recombination of electron-hole pairs, it is also known as radiative recombination. 
Luminescent rates are highly material dependent and are high in doped or direct-bandgap semiconductors, and in materials with deep-trap impurities.  Phonons can also be generated as electrons deexcite. In indirect-bandgap semiconductors, both relaxation to the bottom of the conduction band and electron-hole recombination occur primarily by phonon emission. The same holds for conductive materials, where electrons can effectively emit heat by relaxing within the conduction band. Future DM experiments will require detector-specific material measurements at relevant operating temperatures to enable modeling and mitigation of this background.

Secondary emission is already an important background consideration in several low-mass DM experiments, including Cherenkov and luminescence photons from tracks passing through detector components~\cite{SENSEI:2020dpa,SuperCDMS:2020ymb,Du:2020ldo,Chiles:2021gxk,SuperCDMS:2020aus}. Proposed and planned experiments may also be affected by secondary emission, such as from phonon emission from electronic excitations in detector holders or from luminescence or Cherenkov emission from near-detector insulators (such as plastics)~\cite{Du:2020ldo,SuperCDMS:2016wui}.  Secondary emission may also be an important background for CCD-based neutrino detectors~\cite{CONNIE:2019swq} and may  be relevant for explaining the loss of coherence of quantum qubits~\cite{Du:2020ldo,Vepsalainen:2020trd}.

\subsubsection{Thermal processes}
\label{sssec:thermal}
\textbf{\textit{Lead:} C.\ Savarese}
\\

Thermal processes in a detector medium may become a source of backgrounds as the sensitivity of DM detectors improves. For noble-liquid TPCs, several experiments have reported observation of ``electron trains'' following intense ionization events~\cite{xenon10LDM,lux2020electrons}. As discussed in \refsec{sssec:spurious_electrons}, a fraction of such events may be due to delayed emission of surface-trapped electrons at the liquid-gas interface. The model in \refcite{sorensen2017electron} fits experimental data on electron-extraction efficiency in Xe and predicts trapping times for thermalized electrons---in the range of \SIrange{6}{23}{\milli\second} under certain conditions---compatible with observations. There remains uncertainty in model parameters, which suggests the need for further research to better characterize this thermal process.

Dirac materials are promising for exploration of DM parameter space in the \si{\keV}--\si{MeV} mass range~\cite{diracmaterials2018LDM}. Their extremely narrow bandgap enables sub-\si{\eV} thresholds, and lattice anisotropy can be exploited to search for daily modulation effects to suppress backgrounds. Detector development is in the R\&D phase, likely leading to avalanche photodiodes. A key background for these sensors will be dark counts. The characterization and mitigation of such noise will play an instrumental role in advancing this novel technology.

Thermal processes in sensors instrumenting physics detectors are also a relevant source of noise that may constrain experimental sensitivity. Dual-phase TPCs are instrumented with high-sensitivity photosensors---photomultiplier tubes (PMTs) or silicon photomultipliers (SiPMs)---that are affected by dark noise. The dark-count rate exponentially decreases with temperature, thus rooting the origin of this noise in thermal agitation. A great deal of R\&D has been carried out to suppress dark noise as much as possible. A second mechanism (independent of temperature) was eventually discovered in SiPMs, possibly related to field-assisted band-to-band tunneling, and may be dominant at temperatures below \SI{100}{\kelvin}. For detectors instrumented with phonon sensors it is important to minimize readout noise, which has led to optimization of SQUID amplifiers. Thermal and related noise in the ancillary systems of next-generation DM detectors will need to be carefully considered.

\subsubsection{Release of accumulated energy in materials}
\label{sssec:accumulated}
\textbf{\textit{Lead:} S.\ Pereverzev}
\\

Recent measurements with low-threshold DM and CE$\nu$NS detectors have reported excess low-energy backgrounds~\cite{Proceedings:2022hmu}. A universal background-production mechanism may explain (some or all of) these results: energy accumulation within detector materials and delayed releases \cite{nygren,Pereverzev:2021zoi}. Interactions among energy-bearing states, defects, and structures can lead to correlations in energy-release processes and avalanche relaxation events. These events can mimic low-energy particle interactions. In self-organized criticality theory, interactions can lead to avalanche-like releases of energy and even catastrophic relaxation events; whereas the tunneling two-level systems (TLS) model excludes these interactions (and avalanches). Such effects are not sufficiently understood and require attention. In some cases, like noble-liquid TPCs and superconducting nanowires, analyses of energy deposition and release mechanisms already allow a path to further decrease unwanted backgrounds and increase energy sensitivity \cite{Pereverzev:2021zoi,Pereverzev:2022bbf}. 

Analyzing excess low-energy backgrounds in rare-event experiments requires considering several material effects that can lead to energy accumulation and delayed releases:

\begin{itemize}
\item	Residual and cosmogenic radioactivity accumulates excess energy in materials: thermally stimulated luminescence, electron emission, and conductivity (TSL, TSEE, TSC) are known effects \cite{chen1997theory}. Mechanical stress and deformation also result in stored energy. Energy releases can occur at different time scales without heating, e.g., delayed luminescence and electron emission in solids, or after-luminescence in gases. Mechanical stress conversions into phase and chemical transformations \cite{ZHANG20011985}, delayed luminescence, and electron emission are known in metals, dielectrics, and semiconductors  \cite{oster,voss,BAXTER1972571}. Numerous examples indicate that these processes are typical for some materials. A better understanding of relaxation processes in systems with energy flow is needed (and non-equilibrium thermodynamics in general) \cite{Pereverzev:2021zoi}. 
\item	In glasses (disordered solids), material flow is non-linear; relaxation processes are non-exponential, lengthy, and hysteretic. Glasses are out of thermal, mechanical, and chemical-structure equilibrium after cooling below the glass transition temperature, and can store excess energy in different forms (defects in crystals, spin interactions in spin glass, etc.). Energy storage and release in glasses therefore represents an important research direction for improving our understanding of excess noise in detectors. The current theoretical understanding of glasses is likely insufficient, particularly with respect to treating interactions among energy-bearing states \cite{BAXTER1972571,bradby}. 
\item	At low temperatures (100 mK and below), relaxation in some material subsystems demonstrates glass-like properties:  deformations and defect motions in crystals \cite{ZHANG20011985,PhysRevLett.76.3136,leggett};  magnetic moments, localized electrons, and nuclear-spin re-orientations, including in metals and superconductors (see, e.g., Refs.~\cite{KORTE20111853,PhysRevB.66.064526}); and electric charge relaxation in dielectrics and semiconductors, both in the bulk and on surfaces/electrodes \cite{PhysRevLett.103.197001,PhysRevLett.113.256801}. Disequilibrium and excess energy accumulation are expected due to ionizing radiation, thermo-mechanical stress, and changing electric and magnetic fields. Interactions among these subsystems make energy-conversion and -relaxation events possible (see Ref.~\cite{Pereverzev:2021zoi} for a short discussion).
\item	The interaction of quantum devices with non-equilibrium excitations remains a challenging problem. Theoretical models need to include quantum tunneling, entanglement, squeezing, and dynamical effects due to interactions between excitations and defects. The absence of such theoretical models should not prevent experimental studies beyond the currently prevailing TLS model \cite{Phillips_1987}. Studying excess background in DM and CE$\nu$NS experiments from a condensed matter and non-equilibrium thermodynamics perspective can potentially provide the ``smoking-gun evidence" that the TLS model needs to be reconsidered  \cite{PhysRevLett.76.3136}. 

\end{itemize}

Non-equilibrium thermodynamics and related material issues appear in many fields, including nuclear and high-energy physics, condensed matter physics, and quantum information science. A research program coordinated across these fields would be beneficial (see, e.g., Ref.~\cite{Pereverzev:2022bbf}).

%% file: Sections/4.1-Particle_transport.tex
 \subsection{Particle Transport}
\label{ssec:particle_transport}
Several codes are commonly used for simulating particles' interactions with detectors.
These codes typically require model development and data, both as inputs and for validation of these models.
While this section focuses on \geant, \fluka, and \srim, their needs are generally representative of those of the wider set of particle transport codes.

\subsubsection{\geant}\label{sssec:particle_transport:geant4}

\geant\ is a Monte Carlo framework for simulating particle interactions with matter. 
Users describe a geometry and choose physics models, and \geant\ propagates primary particles through this geometry.
It was developed for accelerator-based applications, but it has since been adopted by other fields, including astroparticle physics.
In astroparticle physics, \geant\ is generally used for conceptual detector design and signal/background modeling.

Electromagnetic physics models have been validated above \SI{1}{\keV}, and the \livermore\ and \penelope\ lists extend it to \SI{\sim250}{\eV}, below which atomic effects limit their accuracy. 
Neutron transport comparisons between \geant\ and \mcnpx\ show broad agreement, though tension arises between them and data in some cases~\cite{hechtComparisonGeant4MCNP62014}.
Comparisons like \refcite{hechtComparisonGeant4MCNP62014} and improved models will be valuable.
Nuclear models also need to be validated and improved, including those for inelastic interactions.
More data are needed for neutron interactions (\eg\ fully-correlated de-excitation cascades following neutron captures), which require exclusive cross-section measurements that are particularly sparse.

Models describing neutron production and cosmic-ray activation calculations also need to be improved.
Comparisons to underground cosmogenic neutron production vary: measurements at Boulby are \SIrange{2}{3}{\times} smaller than \geant's predictions~\cite{araujoMeasurementsNeutronsProduced2008}, while KamLAND found mixed results comparing cosmogenic isotope production~\cite{kamlandcollaborationProductionRadioactiveIsotopes2010}.
Needs for improving radiogenic neutron production calculations are discussed in \refsec{ssec:radiogenic_neutrons}, cosmogenic neutron production in \refsec{ssec:cosmogenic_neutrons}, and activation calculations in \refsec{sssec:bulk_contaminants:cosmogenic_isotopes}.
As these needs are addressed, their results should be integrated into \geant\ and other particle transport codes.
Continued development of \geant\ is therefore important, as are measurements to test it.

\geant's electromagnetic physics is generally adequate for current applications, though improvements below \SI{250}{\eV} and down to sub-\si{\eV} energies will be needed for future low-threshold detectors.
Neutron physics exhibits larger uncertainties: several-fold inconsistencies are reported for neutron transport comparisons with data in \refcite{armengaudBackgroundStudiesEDELWEISS2013}.
Thermal neutron transport agrees between \mcnp\ and \geant\ and measurements at the few-percent level where data are available~\cite{VANDERENDE201640}, but larger discrepancies arise in other situations. 
Strong agreement is achieved by explicitly including thermal scattering cross sections in \geant's models where data are available, though agreement may break down at cryogenic temperatures where thermal energies are lower, and thermal cross-section measurements are only available for a limited number of isotopes.
Additional thermal neutron scattering data for more materials would be highly beneficial as it would allow for improved accuracy in modelling low-energy neutron physics in detector materials, which can in turn improve uncertainties on neutron background models for various DM and neutrino experiments.

Uncertainties in \geant's transportation and production of detectable quanta (\eg\ photons, electrons, phonons, \etc) limit its accuracy modeling detector responses~\cite{brandtSemiconductorPhononCharge2014}, especially as more multivariate analyses demand more precise microphysics models.
Uncertainties in atomic data, including atomic relaxation and interactions, often hinder such uses of \geant.
Improved physics models for tracking detectable quanta, including models for creating processes (\eg\ electron-ion recombination) would be useful for future experiments.
Accurate modeling of atomic physics, including relaxation processes with \xr\ and Auger and conversion electron emissions~\cite{incertiSimulationAugerElectron2016,abdurashitovResponseProportionalCounter2016}, is also important as experiments push to lower thresholds, with \si{\eV}--\si{\keV} signals becoming increasingly critical~\cite{barkerLowEnergyBackground2016,Li_2021}.

\subsubsection{\fluka}\label{sssec:particle_transport:fluka}
The \fluka\ software is separately managed by CERN~\cite{FLUKA_1,ahdida2022new} and INFN~\cite{bohlen2014fluka,ferrari2005fluka} as of 2019. 
It is widely used to evaluate cosmic-ray muon-induced backgrounds, including neutron and isotope production. 
As both forks  evolve, distinct names would avoid confusion.

Neutron physics in \fluka\ derives from models and data evaluations.
Limited data $\lesssim$\SI{10}{\MeV} requires uncertain models to be relied upon, which more measurements could improve. 
Data and models can disagree by \SI{10}{\times} in targets with anti-resonances like \ce{^{40}Ar}~\cite{wintersTotalCrossSection1991}.
CAPTAIN has made several measurements~\cite{bianCAPTAINExperiment2015}, and more would help.
Above \SI{100}{\MeV}, neutron data are limited by the difficulty of producing such high-energy beams, though proton data up to \SI{\sim1}{\TeV} would help.
Measurements in \ce{Ar} are particularly needed, as is testing inconsistent evaluations and quantifying uncertainties in \fluka's calculations.

Photonuclear data needed to simulate muon interactions are very sparse. 
Muon and photon beam measurements would help, especially on common targets like \ce{Ar}.
Measurements on \ce{Ti} at Jefferson Lab show good agreement for \gamman, though \gammap\ shows tension with models.
While \ce{C} is well-studied, data on other targets will improve models.
Recent evaluations focus on inclusive cross sections, and exclusive cross-section data are particularly lacking, though they are needed to simulate correlations among emissions.

\ENDF\ evaluations are model-dependent $\gtrsim$\SI{10}{\MeV}, leading to \SI{>30}{\percent} uncertainties; data and evaluations bringing these uncertainties to \SI{\sim10}{\percent} are needed. 
More \ngamma\ measurements on common materials would also help. 
\ce{Ar} and \ce{C} are well-measured, though \ce{Xe} and \ce{Gd} are not. 
Particularly, the correlations between emitted \gammars\ are not well known. 
For example, de-excitation $\gamma$-ray branching ratios in \ce{Gd}\ngamma\ are known to \SI{18}{\percent}, which obfuscates correlations within that, including the subsequent de-excitation pathways.

Cosmic-ray measurements provide valuable high-energy data.
The field would benefit from more interactions with the cosmic-ray community; measurement from AMS have been useful for building \fluka's models and for understanding cosmic-ray physics.
Because these interactions depend on solar, atmospheric, and geomagnetic conditions, models for translating measurements between locations would help, as would opening communication with the community mapping Earth's magnetic field evolution.
These factors would improve background models for cosmogenic neutrons (\refsec{ssec:cosmogenic_neutrons}), cosmogenic activation (\refsec{sssec:bulk_contaminants:cosmogenic_isotopes}), and atmospheric neutrinos (\refsec{ssec:astrophysical_neutrinos}).
The most widely-used model of the atmospheric neutrino flux---which is highly uncertain and varies with location and time---comes from \fluka~\cite{Battistoni:2005pd}. 
Improving these models will therefore help as experiments approach the neutrino floor.
It may be possible to constrain the flux with \textit{in-situ} muon flux measurements, though doing so requires knowledge of the time-varying muon energy.
Near-surface muon spectral monitors may be helpful to this end. 
Improving atmospheric neutrino models requires a better understanding of $\pi/K$ production in cosmic-ray showers up to $\mathcal{O}(\SI{10}{\GeV})$, which may benefit from NA-61.
Some constraints can be placed by underground experiments at varying depths measuring muon flux modulation, particularly by correlating the flux with local atmospheric conditions, as done in \refscite{anSeasonalVariationUnderground2018,belliniCosmicmuonFluxAnnual2012,abrahaoCosmicmuonCharacterizationAnnual2017,grashornAtmosphericChargedKaon2010,bouchtaSeasonalVariationMuon1999,andreyevSeasonBehaviourAmplitude1991,barrettAtmosphericTemperatureEffect1954,desiatiSeasonalVariationsHigh2011,minoscollaborationObservationMuonIntensity2014,minoscollaborationObservationMuonIntensity2010,shermanAtmosphericTemperatureEffect1954,cutler1981meteorological,collaborationFluxModulationsSeen2016}.

In general, more work developing \fluka\ is needed. 
Help would be beneficial for improving both interaction models above and below \SI{100}{\GeV}, modeling electron-induced deep-inelastic data, benchmarking code, and generally improving models.

\subsubsection{\srim/\trim}\label{sssec:particle_transport:srim}
Stopping and Range of Ions in Matter (\srim) is a Monte Carlo package that calculates several parameters for ions interacting with matter between \SI{10}{\eV} and \SI{2}{\GeV}~\cite{ziegler1988stopping,ziegler2013stopping}. 
The core of \srim\ is the \trim\ program (Transport of Ions in Matter), which handles the final distribution of ions and the kinetic effects associated with their energy loss, such as target damage, sputtering, ionization and phonon production~\cite{biersack1980monte}.  
Since the 1980s when it was introduced, \srim\ has been widely used in the ion-beam community where it is highly recognized as a useful and well tested tool for ion implantation research and radiation material science. In the DM community, the usage of \srim\ has increased over the years because \geant---originally geared towards high-energy accelerator physics---was not well-suited for simulating low-energy (${\sim}$\si{\keV}) ion interactions until recently~\cite {mendenhall2005algorithm}. \srim\ can be used, e.g., to model under a simplified geometry the implantation profile of \ce{^222Rn} progenies that have plated out onto detector surfaces as discussed in \refsec{ssec:surface_contaminants}. It has also been used to model with precision the energy loss processes of these progenies (including for the \ce{^206Pb} ion at \SI{103}{\keV}), as well as their resulting recoil spectra at the detector surface~\cite{angloher2012results,kuzniak2012surface}. \srim\ has also been used to validate extensions of \geant\ that were developed to model low-energy particle interactions~\cite{mendenhall2005algorithm} so as to have a more general framework in which these interactions can be handled in naturally complex and arbitrary detector geometries.

\srim\ can output the distribution of final-state energy dissipated through phonons and ions, allowing for the calculation of final-state energy partition seen as heat production (\eg\ signals in bolometers or bubble chambers) or scintillation and ionization modes. As a result, \srim\ can also be used as part of quenching factor calculations and for simulating surface backgrounds.
However, limitations in \srim's accuracy for slow ions limit its precision for these purposes. The accuracy of \srim\ (as of 2010) is discussed in \refcite{zieglerSRIMStoppingRange2010a}, and a detailed discussion of the input data and model accuracy (from 1985) is given in \refcite{zieglerStoppingRangeIons1985}. In particular, this paper summarizes the accuracy at various ion energies $E$ as,
\begin{itemize}  \setlength\itemsep{0em}
    \item $E > \SI{10}{\MeV\per\amu}$: Well-developed theoretical models; accurate within \SI{5}{\percent},
    \item $E > \SI{1}{\MeV\per\amu}$ : Reasonable theory and scaling laws; accurate within \SI{10}{\percent},
    \item $E > \SI{0.2}{\MeV\per\amu}$ : Theory and scaling laws supported by data; accurate within \SI{20}{\percent},
    \item $E<\SI{0.2}{\MeV\per\amu}$ : Theory only, with no data beyond \ce{H} and \ce{He}; accurate to \SI{\sim2}{\times}.
\end{itemize}
Notably, recoiling nuclei like \ce{^206Pb}---important for surface and dust backgrounds, where their unattenuated energy is \SI{0.5}{\keV\per\amu}---fall into the least accurate energy range. 
While more recent versions of \srim\ have improved calculations, \refcite{zieglerSRIMStoppingRange2010a} reports \SI{200}{\percent} disagreement for \ce{Mg} ions below \SI{100}{\MeV\per\amu}. 
Ion velocities tend to change rapidly for energies between \SI{10} and \SI{100}{\keV\per\amu}. Additionally, the underlying theory used for these ions draws from that developed by Lindhard, Scharff, and Schiott~\cite{lindhardIntegralEquationsGoverning1963}, which loses validity for $\epsilon\lesssim\num{e-2}$, where $\epsilon$ is the medium-dependent reduced energy variable defined in \refcite{lindhardIntegralEquationsGoverning1963}.
Recoiling \ce{^206Pb} nuclei fall near or below this value, making \srim\ less reliable.
\refcite{sorensenAtomicLimitsSearch2015a} also points out that atomic effects begin to play an important role near these energies.
Indeed, \refcite{xuFirstMeasurementSurface2017} found that recoiling \ce{^210Pb} nuclei in LAr are quenched more strongly than predicted by \srim. Similar challenges are reported more broadly in \cite{hitachiLuminescenceResponseQuenching2021} for LAr and LXe, which identifies a need for more quenching factor measurements of low-energy heavy ions, particular for \ce{Pb}.
Improving models of backgrounds involving such slow nuclei, such as backgrounds arising from (possibly attenuated) $\alpha$-decays on detector surfaces or in particulates, requires improving these low-energy stopping power calculations and additional measurements of these ions' scintillation and slowing-down behaviors.

%% file: Sections/4.2-Detector_response.tex
 \subsection{Detector Response}
\label{ssec:detector_response}

Geant4 was developed for high-energy particle physics 
and is still closely tied to its HEP roots. Tracks are treated as classical free particles with mass, energy, and direction; the momentum is computed from $p^2 = E^2 - m^2$.  ``Low energy'' interactions are ionization and energy-loss related. 
While this framework is quite suitable for simulating background and signal interactions in DM searches, it is not immediately suitable for simulating the detector response to those backgrounds and signals.  Liquid-noble detectors have developed the NEST library~\cite{nest} on top of Geant4 for performing detailed simulations of the complex scintillation light response in their materials.  Solid-state detectors also require a detailed detector-response simulation such as with the G4CMP module in Geant4~\cite{Leman,g4cmp}, covering the production and transport of charge carriers and phonons in crystals. 

\subsubsection{Liquid-noble detectors}
\label{sssec:NEST}
The Noble Element Simulation Technique (NEST) is available specifically for DM direct detection in Xe~\cite{nest}, at both zero and nonzero electric fields, for TPC and non-TPC technologies, and for single- and dual-phase detectors. It has also been used for Ar-based detectors; however, other models and software packages are also available: LArSoft~\cite{Church,larsoft} and LArQL~\cite{Marinho:2022xqk} for neutrino physics, and PARIS~\cite{DarkSide:2017wdu} and RAT~\cite{rat} for DM. NEST has international members of multiple experiments and seeks funding and personnel as its own standalone collaboration to achieve an interdisciplinary unification for the DM and neutrino communities that addresses modeling challenges common to both. 
NEST can rapidly convert raw energy depositions into final quanta of photons and electrons, including production of scintillation pulse shapes as a function of a variety of parameters.  
NEST delivers not only expected values but also distribution widths and rare tails, 
and it can assist in modeling thresholds and efficiencies precisely as smooth functions.
 
Where NEST struggles the most, having been constructed from semi-empirical models, is robust sub-keV extrapolation, in particular for NRs whose calibration challenges are addressed in Sec.~\ref{ssec:NR-calibration}. Lack of data, especially on primary photon yield, and lack of good modeling limit NEST's predictive power in the sub-keV regime, where NEST suffers from systematic uncertainties in the underpinning calibration data and from there being an incomplete understanding of near-threshold phenomena. In turn, progress is hampered by lack of funding for dedicated calibrations.
Another issue is a paucity of data for the lower fields in some of the larger experiments; most of the robust calibration data sets were taken at much higher fields. Addressing these challenges will require creative new calibration sources, as well as support for tackling noble-element microphysics from first principles to deal with energy regimes far removed from traditional HEP and more akin to atomic physics and quantum chemistry. Funding centralized efforts in this direction may enable more efficient use of the limited resources available for this type of work.

\subsubsection{Solid-state detectors}
\label{sssec:G4CMP}

The ability to integrate particle physics and solid-state detector response into a single simulation is appealing for several reasons: identical geometries; use of existing libraries for material properties, particle transport, 
and data collection and reporting; and integration of simulation data into a single output. 
G4CMP~\cite{g4cmp} addresses these needs by integrating the anisotropic behaviour of charge carriers in semiconductors into the Geant4 framework. 
Properties of the crystal lattice are associated with volumes using the same method as Geant4's ``field handlers'' and interactions of phonons and charge carriers are implemented as Geant4 ``processes,'' with the Geant4 toolkit handling the selection of processes, transport between interaction steps, and modification of particle kinematics. A unique feature of G4CMP is the anisotropic transport. For phonons, this is handled via the difference between phase and group velocities, and a vector mapping between them. For electrons, G4CMP uses a 3$\times$3 tensor to represent the electron's effective mass in different directions near the minimum of the energy band and to compute momentum and energy.
This reproduces the behaviour of electrons when transported under an electric field, which follow ``valleys'' in physical space. 

G4CMP is a step in the right direction. However, the existing simulation tools are still relatively limited, having been designed for indirect-gap materials and for sources at the eV scale and higher. Additional processes that occur in other types of materials need to be incorporated (e.g., scintillation in direct-gap semiconductors). Enabling use of novel materials as bulk targets---polar materials, magnetically ordered materials, superconductors---will increase the utility of these tools for a large class of new HEP experiments and QIS-related applications. 
Further, while some of the basic phonon-quasiparticle processes have been implemented in G4CMP, more advanced treatment in thin films is necessary to better model the complex response of superconducting sensors.   
In many cases, this will require tracking energies to sub-meV-scale precision 
and simulation of quasiparticle diffusion. Initial work to this effect has been done~\cite{Hochberg:2021ymx}, but not to the level of the particle tracking implemented in G4CMP. Such simulation is needed to understand the complex responses of MKIDs, nanowires, TESs, and Josephson-junction-based sensors for a wide range of particle interactions, and to study quasiparticle-poisoning effects in these devices. Development in this area is ongoing, but the scale of the current effort is not keeping pace with the growing need in the HEP and QIS communities.

%% file: Sections/5.1-Radioassasys.tex
\subsection{Radioassays}
\label{ssec:radioassays}
\textbf{\textit{Lead:} M.L.\ di Vacri}\\
\textit{Contributors: A.\ Piepke}
\\

Next-generation DM and $0\nu\beta\beta$-decay searches need to achieve unprecedentedly low backgrounds to improve upon existing constraints. These extraordinarily tight background requirements necessitate that materials used in the construction of such experiments be screened to assess their levels of radioactivity. This must be done well before start of construction to assure that only suitably low background materials are used; otherwise, experiments may face high backgrounds and fail to meet their sensitivity goals. Access to suitable screening methods and facilities is therefore a prerequisite for scientific success.

With respect to natural radioactivity, \ce{^40K}, \ce{^232Th}, and $^{238}$U are among the largest concerns. These radionuclides are found virtually everywhere and mixed into everything. For the most demanding components near the active detector medium, next-generation experiments typically require assay sensitivity to $\mathcal{O}$(\SI{}{\micro\becquerel\per\kg}) activities or below. For comparison, in natural Xe 2$\nu\beta\beta$-decay corresponds to an activity of about \SI{4}{\micro\becquerel\per\kg}. \ce{^232Th} and \ce{^238U} radioactivity contribute via decay chains to the detector background, an analytical complication because it is often the decays of the progeny further down the chain that create background. Knowledge of the existence of decay-chain equilibrium or disequilibrium is often required to convert radioactivity measurements into background predictions.

Analysis methods generally fall into two categories depending on the decay half-lives: counting decays \textit{vs.}\ counting atoms.
Detection of decay radiation provides the most direct and assumption-free input to background models. In the case of the natural decay series, low-background counting facilities typically detect the $\gamma$ radiation resulting from the relatively short-lived decays of \ce{^214Pb} and \ce{^214Bi} in the \ce{^238U} series, and \ce{^228Ac}, \ce{^212Pb}, \ce{^212Bi}, \ce{^208Tl} in the \ce{^232Th} series.  These can constrain the important \ce{^228Ra} and \ce{^226Ra} activities.
However, while conceptually clean, measurement sensitivities are typically limited to tens of \si{\micro\becquerel\per\kg} due to small decay rates, even with large counting exposures. Absorption of the emitted $\gamma$ radiation by the sample itself and limited detection efficiencies render this method impractical for the most demanding samples.

While the long lifetimes of \ce{^40K}, \ce{^232Th}, and \ce{^238U} result in low decay rates, they also result in relatively large populations in terms of the number of atoms present in a sample. Counting the number of meta-stable atoms therefore offers an alternative to counting decays. Inductively Coupled Plasma Mass Spectroscopy (ICP-MS) and Neutron Activation Analysis (NAA) are commonly utilized approaches for the detection of these isotopes in the \si{\ppt} to sub-\si{\ppt} range---concentrations resulting in $\mathcal{O}($\SI{}{\micro\becquerel\per\kg}) activities. However, while offering clear advantages in measurement sensitivity, both ICP-MS and NAA determine the long-lived heads of the \ce{^232Th} and \ce{^238U} decay chains and not their background-producing progeny. 
Most experiments have had to assume decay-chain equilibrium, justified by successful predictions of background rates. There is no known analytical method capable of verifying this assumption to the level of sensitivity needed. The assumption of chain equilibrium, paired with a case-by-case evaluation of the chemistry utilized in the making of each material, might be considered an irreducible risk.

Many experiments have implemented exhaustive screening programs combining different techniques (see, \eg,  \refscite{rad_exo-200_2008,Cebrian_2015,ABGRALL201622,pandaXrpurity,Cebrian_2017,rad_exo-200_2017,GERDArpurity,XENONrpurity,lz_assay_2020}). These techniques are described below. 

Whilst the field of radioassay is well-developed, future experiments would benefit greatly from specialised facilities where materials can be screened using many techniques in the same physical location; for example a facility housing Ge detectors, ICP-MS, Rn emanation, and $\alpha$ screening. This would allow for a single material to have its radioactive contamination fully mapped out, including fully characterising all parts of a decay chain, without the risks of contamination or loss that come with transport and handling, or the complication of systematic errors from different analyses with different software. Greater precision across each technique will also be needed, and due to the higher and higher sensitivities of future experiments, increasing throughput is important so that the capability to screen items for longer amounts of time exists. Establishing a more streamlined route to tying together radioassay measurements into physics backgrounds is also important.

\subsubsection{High Purity Germanium (HPGe)}
\textit{Contributors: S.\ Cebrian, S.\ Scorza, C.\ Ghag, B.\ Mount, M.\ Laubenstein}
\\

\gammar\ spectrometry performed underground using HPGe detectors is extensively used to screen materials. Characteristic radiation 
is used to quantify radionuclides of interest. This method can inform on a possible disequilibrium in the naturally occurring decay chains, as the activity of different sub-chains is independently determined. Being a non-destructive method, it is possible to assay the components actually used in an experiment. Any sample matrix sample---solid, liquid, powder---can be measured and no particular sample preparation other than external cleaning is necessary. Large samples (hundreds of grams) and long measurements over weeks are typically required to achieve high sensitivity.

It is estimated that around 90 Ge detectors are in operation all over the world. Special facilities, some even at shallow depth, have been developed \cite{KOHLER2009736,kimballton,giove}. Commercial detectors can measure activities at the level of \SI{\sim1}{\milli\becquerel\per\kg}. To increase sensitivity down to $\mathcal{O}$(\si{\micro\becquerel\per\kg}), custom-made detectors have been developed to achieve  ultra-low background conditions, using selected radiopure materials and shielded electronics \cite{gempi,gator}. Other efforts to improve performance are devoted to increase detector efficiency \cite{obelix,cage}. 
Maintaining a wide range of Ge detectors to meet the diverse material sample types and specific isotopic sensitivities required for low-background experiments is essential. Types of Ge detector include large-mass and high-efficiency p-type HPGe, low-threshold planar Broad Energy Germanium (BEGe), and well-type high resolution Small Anode Germanium (SAGe)---well-types offer $4\pi$ counting. 

Facilities for \ce{Ge} screening include: the Boulby Underground Germanium Suite (BUGS) with 3 p-type coaxial detectors, 2 planar BEGe detectors, and 1 well-type SAGe detector; Canfranc Underground Laborory with 7 p-type coaxial detectors; Gran Sasso National Laboratory with 12 p-type coaxial detectors, 1 well-type detector, 1 n-type closed-end coaxial detector, and 1 point contact BEGe detector; SNOLAB with 5 p-type coaxial detectors; and Black Hills Underground Campus (BHUC) with 4 p-type and 1 n-type coaxial detectors. 

\subsubsection{Neutron Activation Analysis (NAA)}
\textit{Contributors: A.\ Piepke}
\\

NAA is based on transmuting long-lived radionuclides, such as \ce{^40K} ($\rm T_{1/2}=\SI{1.25e9}{\year}$, \ce{^232Th} ($\rm T_{1/2}=\SI{1.40e10}{\year}$), and \ce{^238U} ($\rm T_{1/2}= \SI{4.47e9}{\year}$), into short-lived species, resulting in higher specific activities and then detecting their decay. This is done by exposing the material to a high neutron flux where the resulting nuclear capture reactions create the desired short-lived species. Using natural radioactivity as an example, the resulting species are \ce{^42K} ($\rm T_{1/2}= \SI{12.355}{\hour}$), \ce{^233Pa} ($\rm T_{1/2}= \SI{26.975}{\day}$), and \ce{^239Np} ($\rm T_{1/2}=\SI{2.356}{\day}$), showing dramatically reduced half-lives when compared to their parents. If the sample is exposed to a sufficiently high neutron flux, a large enough fraction of target atoms is transmuted to achieve a boost in the observable activity. This is the reason NAA is often performed using research reactors, offering sufficiently high neutron fluxes to create substantial product activities. This scheme also works for other stable or meta-stable target nuclides of interest.
Counting of the activated samples is often done using \ce{Ge} spectroscopy. Using long activation times at a high flux reactor and counting with a single Ge detector, one can routinely achieve \si{\ppt} or even sub-\si{\ppt} sensitivity for \ce{^232Th} and \ce{^238U}. \ce{K}-sensitivities below \SI{1}{\ppb} are also routinely obtained. Sensitivity can be enhanced  to the level of \SI{0.01}{\ppt} of \ce{U} and \ce{Th} by techniques such as $\gamma$-$\gamma$ coincidence counting and post-activation separation of nuclides from background-creating chemical impurities.  

The LZ and nEXO research groups at the University of Alabama (UA) have a long-established NAA capability. Small samples are typically activated at the MIT Nuclear Reactor Laboratory (MITR). Pre- and post-activation sample preparations and and sample counting are done at UA. Radioassays have been performed for the KamLAND, EXO-200, nEXO and LZ projects. In a recent study the UA group showed that an order of magnitude sensitivity enhancement in detection of U can be achieved though the utilization of \ce{Ge}-\ce{Ge} coincidence counting~\cite{tsang_2021}. The group plans to set up such a capability in case the required funding can be obtained and there is demand for this technique.

\subsubsection{Inductively Coupled Plasma Mass Spectrometry (ICP-MS)}
\textit{Contributors: C.\ Ghag, J.\ Dobson, M.L.\ di Vacri, S.\ Nisi}
\\

ICP-MS is widely accepted as the fastest and most sensitive analytical technique for ultra-trace elemental analysis. It can measure nearly all the stable isotopes and long-lived radionuclides, and allows for rapid, high-sensitivity determinations of primordial radionuclides \ce{^40K}, \ce{^232Th}, and \ce{^238U} in candidate materials. 

Several groups steward ICP-MS capabilities used by the rare-event search community. The University College London (UCL) ICP-MS system is housed in a dedicated ISO Class 6 cleanroom reserved for assays of materials for low-background experiments \cite{icpms-ucl}. Initially established for the LZ construction project, the facility has since been upgraded to house a state-of-the-art Agilent 8900 tandem mass triple quadrupole ICP-MS, optimised for extremely low detection limits as necessitated by current- and next-generation experiments. 
PNNL has a dedicated facility for handling, preparation, and assay of ultra-low-background materials. Two triple quadrupole ICP-MS instruments (Agilent 8800 and Agilent 8900) are operated with detection limits in the sub-ppt range for \ce{^232Th} and \ce{^238U} determinations in a wide variety of materials; some examples are reported in Refs.~\cite{arnquist2020automated,arnquist2019mass}.
Canfranc Underground Laboratory operates a Thermo Scientific iCAP RQ ICP-MS apparatus, with a mass range from \SIrange{2}{290}{\amu} and detection limits at \SI{1}{\ppt} for \ce{Th} and \SI{0.1}{\ppt} for \ce{U}.
LNGS operates an inorganic mass spectrometry facility. Two ICP-MSs are installed in an ISO 6 class cleanroom: a high-resolution double focusing HR-ICP-MS (Element 2, Thermo Fisher Scientific) and a last-generation single quadrupole ICP-MS with collision/reaction cell (Agilent 7850). Detection limits ranging from \si{ppq} to \si{\ppt} are usually achievable for \ce{Th} and \ce{U}, and limits in the sub-\si{ppq} range were attained for determinations of \ce{^226Ra}~\cite{copia2015low}. 

\subsubsection{Radon}
\textit{Contributors: A.\ Piepke, C.\ Ghag, E.\ Perry, R.\ Bunker, C.\ Jackson}
\\

As described in Sec.~\ref{sec:active_bulk_fluid}, Rn emanation from construction materials presents a major (and in some cases dominant) background to DM searches. 

The most direct mitigation at present is to screen candidate materials for Rn directly~\cite{RadonEmanationRauHeusser,RadonEmanationZuzel,RadonEmanationThesisLiu}. Residual activity from selected materials is also quantified through such assays, informing the experiment’s background model. Radon emanation material assays are typically performed at room temperature. However, this results in large uncertainties when translating screening results to expected rates in noble-liquid targets due to the suppression of Rn diffusion within and subsequent emanation out of materials; so, additional cold measurements are desirable.

In the UK, three facilities for Rn emanation exist; University College London, used for the SuperNEMO and LZ experiments~\cite{lz_assay_2020}, Boulby Underground Laboratory, and the Cryogenic Radon Emanation Facility (CREF) at the Rutherford Appleton Laboratory (RAL), with sensitivities below \SI{0.1}{\milli\becquerel} to \ce{^222Rn}. All three utilize Rn concentration lines and emanation chambers of various sizes instrumented with \ce{Si} PIN diode detectors and were originally developed by UoT. As its name suggests, CREF has vessels that can be cooled down to \SI{77}{\kelvin} for measurement of emanation as a function of temperature. 

In the US, the University of Maryland and South Dakota Mines~\cite{BowlesThesis} have Rn emanation facilities also based on \ce{Si} PIN diode detectors, while the University of Alabama has a system that utilizes the high solubility of Rn in organic liquid scintillator~\cite{lz_assay_2020}. Emanated Rn is accumulated, flushed through liquid scintillator, and then viewed by a PMT to measure \ce{^214Bi}-\ce{^214Po} coincidences to infer \ce{^222Rn} activity. There is also an emanation system at PNNL which detects \ce{^222Rn} with ultra-low-background proportional counters~\cite{WOS:000320282800054}, and a cryogenic emanation capability is under development.

Finally, at SNOLAB, three Rn boards consisting of primary and secondary cold traps are used to concentrate and transfer Rn for counting. Two are underground---a mobile board and a board built into the ultra-pure water system---and one is in the surface clean laboratory . 
The mobile board has numerous ports into which different air volumes can be introduced and can sample from various locations in the laboratory. The board is sensitive to a \SI{0.1}{\milli\becquerel} emanation source. 
 
\subsubsection{Alpha screening}
\textit{Contributors: R.\ Schnee, R.\ Calkins, C.\ Ghag}
\\

Alpha counters measure surface contamination by detecting $\alpha$ particles through a technique such as the ionization of Ar gas. Currently, the most sensitive detector for $\alpha$ screening is the XIA UltraLo-1800~\cite{xia}, with a  sensitivity to surface \ce{^210Po} $<$ \SI{0.1}{\milli\becquerel\per\square\meter}~\cite{BUNKER2020163870}. Operation underground reduces backgrounds from cosmic rays by about a factor of 3~\cite{xinranliuLRT2019XIA}. In-air decays from airborne Rn can be distinguished via pulse-shape discrimination (PSD); so, the dominant background is found to be $\alpha$ decays originating on the tray holding the sample of interest. This background could be mitigated by using Ar that is purified via chromatography~\cite{PUSHKIN2018267} or distillation~\cite{XENON100:2017gsw}.
Further reductions of $\alpha$ backgrounds may be achieved by providing better PSD or improved tracking of events.  Increasing segmentation of the electrodes from 2 to 64 has shown initial promising results~\cite{McNally2017LRT}.  Replacing the electrodes with crossed wires to provide full TPC tracking should essentially eliminate $\alpha$ backgrounds, resulting in a signal-limited sensitivity~\cite{LRT2013bunker}.   

\ce{Si} spectrometers are also used for $\alpha$ screening. An advantage of such devices is the potential of short times between sample exposure and screening, allowing them to be used to assay Rn progeny plate-out using \ce{^214Po}, with greater sensitivity and immediate results compared to the XIA Ultralo-1800's measurements of \ce{^210Po}, due to the \SI{2000}{\times} higher decay rate of \ce{^222Rn} compared to \ce{^210Pb}~\cite{LRT2017MorrisonPlateout}.
\ce{Si} spectrometers may also be operated within large-volume vacuum spaces formed by ultraclean materials, allowing surface-$\alpha$ assays of materials without the XIA constraint that they be flat~\cite{ReichenbacherSURF}.

\subsubsection{Determinations of \texorpdfstring{$^{210}$}{}Pb in material bulk}
\textit{Contributors: R.\ Calkins, E.\ Hoppe}
\\

In detector construction materials, \ce{^210Pb} in the \ce{^238U} decay chain is a common contaminant and source of backgrounds. It can enter materials through a variety of processes.
As described in previous sections, \ce{^222Rn} eventually decays to \ce{^210Pb} and can plate out on to surfaces of materials. If that material is reworked, the surface contamination can enter the bulk. 
Similarly, \ce{U}/\ce{Th} in dust can contribute if it enters the bulk during processing such as melting or machining processes. \ce{^210Pb} is also present in lead which is often used as a shielding material or as a component in solder in electronics. 

In addition to surface $\alpha$ screening, bulk assays have been performed using an XIA UltraLo-1800 to assay \ce{^210Pb} in copper~\cite{ABE2018157}. The low background of the XIA makes this an attractive avenue to pursue.  Simulations are need to account for the \alphap's energy loss in the bulk. The experimental difficulty arises from the surface features on the sample which must be small enough to not affect $\alpha$ emission. This sets the characteristic scale of surface features to less than a few \si{\micro\meter}. With these types of assays and techniques, sensitivity down to a few \si{\milli\becquerel\per\kg} is achievable. 
Assuming the sample surface is smooth, this type of assay may be preferable for some samples because it is a non-destructive technique.  
Assays to ascertain levels below a few \si{\milli\becquerel\per\kg} may be possible but will likely be a destructive technique requiring the concentration of \ce{^210Po} or \ce{^210Pb} for subsequent assay. \alphap\ screening will remain the main technique, unless significant strides to improve ICP-MS sensitivity can be accomplished.

\subsubsection{Determination of Kr in Xe}
\label{sec:KrinXe}
\textit{Contributors: D.\ Akerib, C.\ Hall}
\\

As noted in \refsec{sec:active_bulk_fluid}, separation of trace Kr from Xe is critical to background reduction in LXe TPCs because the dissolved $\beta$-emitter \ce{^85Kr} cannot be eliminated with fiducial cuts. If left at naturally occurring levels in research grade Xe, Kr would overwhelm recoil discrimination capabilities. Alongside reduction techniques (described in Sec.~\ref{ssec:material_purity_infrastructure}) are sensitive assays required to monitor gas processing, assay of the final product, and \textit{in-situ} monitoring of operating experiments to ensure that re-contamination through air ingress is minimized. A well established technique used in the LUX and LZ experiments is the cold-trap assisted residual gas analyzer (RGA)~\cite{DOBI20111}. Typical sensitivity of commercial RGA units is about 1 part in \num{e7}, whereas current generation DM searches require natural krypton concentrations of 1 part in \numrange{e13}{e14}. The cold-trap assisted technique exploits the specialized problem of reducing the Xe ``background'' so that trace Kr can be detected at the requisite concentration. By carefully flowing the sample to be assayed through a LN-cooled trap, the Xe partial pressure introduced to the RGA is suppressed to its sublimation pressure. In contrast, trace Kr is only slightly reduced by adhering to the walls of the trap. Through careful tuning and calibration, current-generation implementation of this technique has achieved 1 part in \num{e14} sensitivity. An alternative technique with comparable sensitivity relies on first reducing the Xe fraction in the sample to be assayed using chromatographic separation before introducing the sample to a customized commercial mass spectrometer \cite{LINDEMANN2014krypton}.\looseness=-1

%% file: Sections/5.2-Material_needs.tex
\subsection{Material Needs}
\label{ssec:material_needs}
\textbf{\textit{Leads:} I.\ Arnquist, E.\ Hoppe}
\\

Current- and next-generation detectors require a wide range of materials with a variety of unique specifications. For example, materials of exemplar radiopurity, mechanical, electrical, and/or thermal properties are required to meet detector sensitivity goals. 

The ultra-low background (ULB) nature of rare-event physics detectors necessitates a ``less is more'' approach. Reducing the total background is paramount, which requires balancing design, engineering, and radiopurity requirements to obtain the desired sensitivity reach of the detector.  Moreover, material production plays a vital role in material radiopurity for ULB physics.  Most industrial material purification and processing steps leverage one or more of the following physical or chemical properties to convert the raw commodities to the final product: melting point, boiling point, reduction potential, chemical affinity, and/or solubility. Processing steps leveraging these parameters can fractionate residual trace elemental contaminants in the material, including those of the $^{238}$U decay series. This fractionation gives way to secular disequilibrium if the procedure significantly favors one relatively long-lived isotope over another. 

Relying on assumed secular equilibrium for next-generation experiments is risky as assay sensitivities do not allow for sufficient discernment of radioactivities of the top, middle, and lower part of the $^{238}$U decay series.  The $^{232}$Th decay series does not have the same likelihood for secular disequilibrium because it does not have long-lived progeny isotopes of significantly different chemistry than the actinides to act as an activity reservoir. The ULB community has identified an ever-growing list of problematic materials when it comes to secular disequilibrium; a good example is lead. For ULB physics purposes, it is safe to assume that the raw lead ore is in secular equilibrium in the $^{232}$Th and $^{238}$U decay chains. However, $^{210}$Pb will fractionate through the many metallurgical processes and be retained in the stable lead, while the $^{232}$Th and $^{238}$U are removed. This is why ancient lead is so desirable; the $^{210}$Pb, with a 22-year half-life, has decayed away. 

The materials in a detector can be loosely grouped into the following types:  structural, conditional, and detecting. Structural materials are those that support the active detector media and supporting architecture of the detector.  Typical structural materials include copper, titanium, stainless steel, acrylic, PTFE, etc.  Conditional materials are those that establish the proper environment for detector operation. Some examples include materials to establish proper operating temperature (e.g., refrigerants), but also shielding materials like lead, water, etc.  Detecting materials are those materials that actually detect the rare event signal and are used to transmit the signal for processing.  They include materials like SiPMs, solders, substrates, cables, wires, electronic components, etc.

Structural materials need to be increasingly radiopure due to the large quantity used in many detector systems; a good example is copper, historically chosen due to its favorable electrical and thermal transfer characteristics.  Demand for copper of of ever-greater radiopurity has grown; electroforming copper further purifies commercially available copper and its use is becoming more widespread. Electroforming copper is a relatively slow process; to attain adequate physical properties and purity, development of methods to increase throughput or alternative purification methods are needed.  In addition, copper is difficult to machine.  Further efforts to create radiopure alloys of copper to increase its hardness and strength are being pursued~\cite{Vitale:2021} which will also decrease the amount needed to meet engineering requirements, thus lowering the radiopurity requirement as well. 

Significant effort has gone into finding commercially available polymers with appropriate radiopurity and structural properties~\cite{Grate:2021}. Many of these have been demonstrated to have adequate radiopurity for near-term experiments even when using 3D printing methods. But ultimately for future applications it may be that adequate radiopurity can only be attained by using polymeric powders or pellets obtained commercially so that the final printing, molding, or extrusion can be performed in cleanroom environments.

Of ever-growing concern are the relatively small (by mass) detector materials themselves.  The small electronic components oftentimes require the utmost in performance, while also necessitating extreme radiopurity. Such devices are typically composite devices composed of a vast array of different materials (\textit{e.g.,} metals, insulators, solders, polymers, ceramics). Each material needs to be validated to meet the stringent performance and radiopurity requirements for inclusion, which can be cumbersome due to the high value of the devices and extremely low radiopurity targets (often in the $\mu$Bq/kg range).  Given the small amount of material to work with for assay validation, assessing secular disequilibrium within the small amounts of detector materials can be extremely difficult or impossible with current assay techniques.

As demand  increases, a route to better materials is via assay campaigns with the help of manufacturers to determine the production methods that lead to the least radio-contamination, such as the campaign to find titanium for the LZ experiment~\cite{Akerib:Ti2017}. 

\newpage

%% file: Sections/5.3-Material_purity_infrastructure.tex
\subsection{Material Purity Infrastructure}
\label{ssec:material_purity_infrastructure}
\textbf{\textit{Lead:} J.\ Hall}
\\

A multi-layered approach of infrastructure in surface and underground laboratories is required to meet the background requirements of next-generation experiments. This can be broken down into several areas: underground production plants, cleaning facilities, purification plants, and radon removal systems.

Underground material production plants will allow greater radiopurity of certain materials. For example, $^{39}$Ar is a radioactive isotope created by cosmic-ray interactions with the atmosphere. Underground Ar (UAr) has been shown to have at least 1000 $\times$ less activity due to the lack of this isotope. The need for UAr is broad and includes high energy physics projects (such as DUNE and DarkSide), nuclear physics (e.g., LEGEND-1000) and applied science such as water age-dating~\cite{Alexander:2019uar}. The main supply of UAr is from a CO$_2$ well in Colorado. An expansion of the Ar extraction plant is underway led by members of the DarkSide Collaboration. This resource is critical to a number of experiments; so, this infrastructure is required to extract the quantities of Ar needed for DM experiments ($\sim$0.5 kT) and any future efforts relying on a significant quantity of this material. Copper, as mentioned previously in Sec.~\ref{ssec:material_needs}, is another material where underground production is beneficial.\looseness=-1 
    
Materials that are near the active volume of rare-event searches often require cleaning before they are incorporated into detector assemblies. Cleaning techniques vary from hand wiping materials with clean water or alcohol to acid leaching to controlled etching of metal surfaces to remove implanted materials. Cleaning removes dust and dirt that build up over time as the materials are machined, wired, and otherwise prepared for experiments. Additionally, radon progeny electrostatically attached to dust can accumulate on surfaces.
    
As noted in Sec.~\ref{sec:active_bulk_fluid} and discussed in Sec.~\ref{sec:KrinXe}, reduction of trace noble-gas $\beta$-emitters from research grade Xe for use in a DM detector is necessary; specifically of concern are $^{85}$Kr and $^{37}$Ar.  
Two well-established methods are gas-phase chromatography \cite{BOLOZDYNYA200750} and LXe distillation \cite{Abe:2008py,Wang:2014,Aprile:2016xhi}. In gas charcoal chromatography, set quantities of Xe are periodically introduced to a column of activated charcoal through which a He carrier gas is flowing. Trace Kr and Ar have weaker binding to the charcoal than Xe and are therefore swept through with a higher average velocity. The system is designed so that most of the trace species exit the column before the Xe and are trapped in LN-cooled charcoal. Once separation is complete, the purified Xe is collected in a cryogenic freezer. Assays are performed at various stages of the process to maintain purity within required parameters using a cold-trap assisted assay described in sec.~\ref{sec:KrinXe}. A 60-kg charcoal column was used by the LUX collaboration to reach 3.5$\times 10^{-12}$ g/g of Kr to Xe \cite{AKERIB201880} and a pair of 400~kg charcoal columns were used by the LZ collaboration to purify more than 10~tonnes of Xe, achieving 4$\times 10^{-14}$ g/g of Kr to Xe and 0.5$ \times 10^{-12}$ g/g of Ar to Xe \cite{AKERIB2020163047,mount2017lzTDR}.
A distillation column features an input condenser, a package tube, a reboiler and a top condenser. Xe is continuously added to the system and is partially liquefied in the input condenser. From there, gas and LXe are fed into the package tube at different heights. The reboiler at the bottom contains a LXe volume that is partially evaporated, while the top condenser liquefies again the up-going Xe gas. In this manner, a counter-flow of up-going gas and down-going liquid is established along the surface of the package tube, so that more volatile gases than Xe, such as Ar and Kr, are enriched at the top and are depleted at the bottom. Thus, ultra-pure Xe can be extracted. This system achieved output concentrations of \isotope[\mathrm{nat}]{Kr}/Xe\,\textless\,\SI{48}{ppq} (\SI{e-15}{mol/mol}) (\SI{90}{\percent} C.L.)~\cite{Aprile:2016xhi}.

%% file: Sections/5.4-Underground_backgrounds.tex
\subsection{Underground lab backgrounds}
\label{ssec:underground_backgrounds}
\textbf{\textit{Lead:} D.\ Woodward}
\subsubsection{Dust and mitigation strategies}
As discussed in Sec.~\ref{ssec:surface_contaminants}, surface contamination during detector assembly and construction contributes to detector backgrounds. For underground locations, dust can be readily generated from exposed rock surfaces, mining, excavation or other construction activities occurring close to the  lab space. Simple mitigation strategies can drastically reduce dust accumulation. Underground facilities utilize clean spaces, which are accessed via transition areas for changing out of clothing that is exposed to the dirty environment. Transition areas can also be defined to transport equipment and detector parts, which are cleaned before entry into the clean space. In addition, air flow patterns can be utilized to prevent dust accumulation and purge sensitive areas. Particle counters are commonly used to assess air quality in such spaces. Additionally, witness coupons and tape lifts may be used during specific assembly procedures to quantify the dust exposure. 

The assembly of sensitive detector volumes is necessarily performed inside of an air-filtered environment (e.g.\ a cleanroom), either at a surface location or underground. After construction, these volumes are sealed before transportation through dusty underground environments. Underground cleanrooms may be permanent or temporary structures, depending on the specific need and duration of the activities. As next-generation DM detectors increase in overall size, construction of cleanrooms of sufficient size may be challenging for some underground locations. 
For current- and next-generation detectors, particularly those that utilize expansive auxiliary systems (e.g., for gas circulation and purification), some components of the overall detector system may be constructed outside of a cleanroom and even outside of dedicated clean spaces. Care must be taken to limit exposure of the internals of these subsystems to the ambient environment to mitigate dust accumulation. This is especially important for those components from which dust (or Rn it emanates) may be transported into a sensitive detector volume (e.g.\ via circulation).

\subsubsection{Radon-reduction systems}
\emph{Contributor: R.\ Schnee}
\\

Various systems may be used to reduce the otherwise high concentration of Rn underground, in order to reduce backgrounds from the fast part of the decay chain, and/or reduce background from Rn progeny plate-out during storage, handling, and assembly of components (see Sec.~\ref{ssec:surface_contaminants}).  For gas volumes that do not have to support human breathing, low-Rn gas from nitrogen bottles or LN boil-off may be used to purge high-Rn air.  Gas bottles result in Rn concentrations typically ranging from  0.5--1\,mBq/m$^3$, while LN boil-off typically is about 0.1\,mBq/m$^3$~\cite{xinranliuLRT2015gases}.  Lower concentrations may be achieved through active filtering of the gas, typically with low-radioactivity carbon columns~\cite{xinranliuLRT2015gases,RadonArSimgen2009}.\looseness=-1

Filtration may also be used to reduce the Rn concentration in the breathable air for cleanrooms used for detector assembly. Systems that reduce the Rn concentration in air by continuous flow through a single filter---typically cooled, activated charcoal---are designed so that most Rn decays before it exits the filter~\cite{nemoLRT2006}. Continuous systems are relatively simple,  are available commercially, and typically achieve reduction factors of $\sim1000$~\cite{ATEKOradonMitigation,mount2017lzTDR}, to $\sim$10--30\,mBq/m$^{3}$.
Alternatively, in a swing system, one stops gas flow well before the Rn has time to exit the filter and regenerates the first filter column while switching flow to a second column~\cite{pocarthesis}.
For an ideal column, no Rn from the input reaches the output, and the only Rn at the output is from emanation from the carbon itself.
Swing systems require less carbon and no refrigeration, reducing costs and space needs. In practice, achieved performance is about the same as that of continuous systems~\cite{LRT2017streetVSA}.  Achieving yet lower Rn concentrations in larger underground volumes would benefit future experiments.  Such improvements may be attainable by simple scaling of previous systems, or by increases in complexity, such as~\cite{LRT2010HallinRadon}.

\subsubsection{Environmental backgrounds from cavern walls}
\emph{Contributors: D.\ Woodward, S.\ Shaw}
\\

At underground locations, the natural radioactivity of the rock is a source of background for direct detection experiments. Naturally occurring, long-lived isotopes such as \ce{^232Th} and \ce{^238U} produce high-energy $\gamma$-rays as well as neutrons via ($\alpha$,n) reactions or from spontaneous fission. Other isotopes such as \ce{^40K}, which undergo single decays, can also contribute to the ambient $\gamma$-ray flux. These particles can produce backgrounds in detectors designed to directly detect DM. 
The common mitigation strategy is shielding of sensitive detector volumes. Neutrons from radioactivity in cavern walls are effectively shielded by water or other low-$A$ materials, while $\gamma$-rays  are shielded by high-$Z$ materials such as steel or lead. In some cases, the shielding itself may need to be carefully selected to be low in radioactivity, especially if it is situated close to the active medium. 

As with other sources of background, particle transport simulations that use accurate geometrical models of the lab environment can be used to assess the impact of these environmental backgrounds. Of particular concern is the possibility that there are poorly shielded regions, for example due to the presence of conduits, pipes, feedthroughs etc. Identifying such regions can help optimize a shielding configuration. In addition, cavern backgrounds may be an important contribution to an overall background model and should therefore be well-characterized prior to data analysis. Typically, these types of simulations are challenging because there is a large flux of particles coming from the cavern, but the probability of a given primary particle to produce a background event in the detector is very small. It is therefore necessary to optimize the transport simulation to reduce computational requirements. A common method is some form of Monte Carlo sample biasing technique, e.g., defining multiple stages of the simulation and multiplying surviving particles at each stage to increase the effective number of primaries that are simulated \cite{akerib2021simulations}. \looseness=-1

In conjunction with simulations of a specific detector arrangement, the $\gamma$-ray and neutron fluxes needs to be understood to properly normalize background rates. One approach is to take samples of the rock that makes up an underground cavern and perform material screening to directly measure concentrations of radioisotopes. Given the availability of assay capability in the community, this is a relatively simple approach; however, underground rock formations can be made up of a number of different rock types and intrusions such that complete coverage of a cavern is difficult. An alternative approach is to directly measure the the $\gamma$-ray flux, which has been successfully done at a number of underground laboratories \cite{akerib2020measurement,malczewski2013gamma,malczewski2012gamma,malczewski2013gamma-2,zeng2014environmental}.

As DM detectors expand in size, the contribution of backgrounds from the underground environment should also increase and must be appropriately quantified and mitigated. Use of additional shielding is one mitigation option, but there may be space considerations for existing underground caverns. The use of veto detectors to tag external neutrons or $\gamma$-rays is another strategy to reduce the contribution of these backgrounds (see Sec.~\ref{sssec:anveto}).\looseness=-1 

%% file: Sections/5.5-Software_infrastructure.tex
\subsection{Software Infrastructure}
\label{ssec:software_infrastructure}
\textbf{\textit{Leads:} C.\ Jackson, S.\ Scorza}
\\

As the background requirements of rare-event search experiments become more stringent and the detectors become larger and more complex, material selection will need significant quality control and quality assurance procedures. Next-generation experiments are likely to require hundreds or even thousands of assays to ensure that components constructed over long periods of time and from different batches of material meet the required standards. Therefore, a robust software infrastructure to track and triage these large-scale assay programs is necessary. This software should:
\begin{itemize}
    \item Act as a record of measurements by individual collaborations, including both selected and rejected materials, components and vendors;
    \item Interface assay results with simulations for background model building, e.g., the Background Explorer framework~\cite{backgroundExplorer} allows rapid analysis of impacts of a material or component choice;
    \item Manage large and distributed assay programs effectively, tracking components across the groups performing the assays; and
    \item Share results across the community, allowing future experimental collaborations to utilize previous experiments' knowledge when choosing materials.
\end{itemize}

Radiopurity.org was developed as a tool to aid tracking and sharing of results~\cite{LOACH20166} across the community. The codebase is available for experimental collaborations to use and a public instance is maintained as a repository of assay measurements. There has recently been an upgrade of this database with improved search options which is now live~\cite{radiopurityUpdate} and there is a growing cross-community effort to develop this software further and increase the number of entries in the public instance. Significant scope exists to improve this tool for the community further, and some examples include tracking additional background sources (e.g.\ Rn emanation), tracking measurement details (e.g.\ energy spectra from HPGe counters), adding more useful sample details (manufacturer, batches, material makeup, etc.), and using the tool to request and manage distributed assay programs.